\documentclass[twocolumn,trackchanges]{aastex701}

\usepackage{graphicx}
\usepackage{xcolor}
\usepackage{natbib}
\usepackage{amsmath}
\usepackage[subfigure]{tocloft}
\usepackage{subfigure}
\usepackage{booktabs}
\usepackage{CJK}
\usepackage{makecell}
\usepackage{multirow}
\usepackage[utf8]{inputenc}

\newcounter{RomanNumber}

\newcommand{\degree}{^\circ}

\received{xxx}
\revised{xxx}
\accepted{xxx}
\shorttitle{Precessing disk in a TDE}
\begin{document}
\begin{CJK*}{UTF8}{gbsn}

\title{Diverse Emission Patterns from Precessing Super-Eddington Disks Formed in Tidal Disruption Events}

\author[0000-0002-3525-791X]{Jin-Hong Chen (陈劲鸿)}
\affiliation{Department of Physics, The University of Hong Kong, Pokfulam Road, Hong Kong, People's Republic of China}
\affiliation{The Hong Kong Institute for Astronomy and Astrophysics, The University of Hong Kong, Hong Kong, China}
\affiliation{Shenzhen Institute of Research and Innovation, The University of Hong Kong, Shenzhen 518057, P. R. China}
\email{chenjh2@hku.hk}
\author[0000-0002-9589-5235]{Lixin Dai (戴丽心)}
\affiliation{Department of Physics, The University of Hong Kong, Pokfulam Road, Hong Kong, People's Republic of China}
\affiliation{The Hong Kong Institute for Astronomy and Astrophysics, The University of Hong Kong, Hong Kong, China}
\email{lixindai@hku.hk}
\author[0009-0008-0526-2604]{Cheuk Kwan Kan}
\affiliation{Department of Physics, The University of Hong Kong, Pokfulam Road, Hong Kong, People's Republic of China}
\affiliation{The Hong Kong Institute for Astronomy and Astrophysics, The University of Hong Kong, Hong Kong, China}
\email{zoekan@connect.hku.hk}
\author[0000-0003-0509-2541]{Tom Man Kwan}
\affiliation{Department of Physics, The University of Hong Kong, Pokfulam Road, Hong Kong, People's Republic of China}
\affiliation{The Hong Kong Institute for Astronomy and Astrophysics, The University of Hong Kong, Hong Kong, China}
\email{kwanman2@connect.hku.hk}
\author[0000-0001-8701-2116]{Zijian Zhang (张子健)}
\affiliation{Department of Physics, The University of Hong Kong, Pokfulam Road, Hong Kong, People's Republic of China}
\affiliation{The Hong Kong Institute for Astronomy and Astrophysics, The University of Hong Kong, Hong Kong, China}
\email{astrozzj@connect.hku.hk}
%\author[0000-0003-4256-7059]{Thomsen Lars Lund}
%\affiliation{Department of Physics, University of Hong Kong, Pokfulam Road, Hong Kong, People's Republic of China}
%\email{lthomsen@hku.hk}

%
\correspondingauthor{Jin-Hong Chen, Lixin Dai}
\email{chenjh2@hku.hk, lixindai@hku.hk}

%%%%%%%%%%%%%%%%%%%%%%%%%%%%%%%%%%%%%%%%%%%%%%%%%%%%%%%%%%%%
\begin{abstract}
A tidal disruption event (TDE) occurs when a star passes within the tidal radius of a supermassive black hole (SMBH). In TDEs it is expected that the orbital angular momentum of the disrupted star is generally misaligned with the SMBH spin axis, which should result in a misaligned super-Eddington disk precessing around the SMBH spin axis due to the Lense-Thirring effect. In this paper, we investigate the distinct observational signatures produced from such TDE disks, by performing radiative transfer calculations upon previous super-Eddington disk simulations. We demonstrate that the precession of the disk and wind drive time-dependent obscuration and reprocessing of X-ray radiation. Depending on the orientation of the viewing angle of the observer and the tilt angle of the disk, four main types of variability are induced: 
1) The smooth-TDEs: The emissions from these TDEs show no fluctuations;
2) The dimmer: The main emission type (X-ray or optical) stays the same, with small to moderate modulations of brightness; 
3) The blinker:  X-ray and optical emissions take turns to dominate in one cycle of precession, with dramatic changes in the X-ray fluxes.
4) The siren: X-ray and optical emissions take over each other twice per cycle, possibly with two different peak X-ray fluxes within one cycle. 
In all three scenarios, we observe an inverse correlation between X-ray and optical emissions. 
Our model provides a physical framework for interpreting TDE multi-wavelength variability through disk precession dynamics and gives an alternative interpretation to the interesting case of J045650.3-203750 which was suggested to be a repeated partial TDE previously. 

%\textit{The dimmer}: when the precession occurs within a narrow range of viewing angles, the light curve experiences only small amplitude of modulations. \textit{Blinker}: Within a moderate range of viewing angles, intense X-ray variability arises as the line of sight transitions from the outflow-cleared polar funnel to the optically thick outflow region, producing an inverse correlation between X-ray and optical/near-UV (NUV) emissions. \textit{The siren}: When the precession spans an extremely wide range of viewing angles, both X-ray and optical/NUV bands show double-peaked light curves over a single precession cycle, reflecting alternating line-of-sight exposure to the inner disk and reprocessing regions. Additionally, for precession periods longer than years, slow disk precession can alter line-of-sight access to the outlfow-cleared polar funel, triggering late-time X-ray flares as the near-polar region rotates into view--a potential explanation for delayed X-ray activity in some TDEs. These results provide a unified framework for interpreting multiwavelength TDE variability through disk precession dynamics. %We apply our precessing disk model to fit the TDE candidate J045650.3-203750, which demonstrates a long-term decay accompanied by superimposed short-term quasi-periodic X-ray variability. Our findings reveal that the precessing disk model can interpret the X-ray and UV light curves of this source with a precession period of approximately 200 days. Additionally, we have constrained the dimensionless black hole spin of this source to be $\gtrsim 0.1$.
\end{abstract}


%%%%%%%%%%%%%%%%%%%%%%%%%%%%%%%%%%%%%%%%%%%%%%%%%%%%%%%
\section{Introduction}
\label{sec:introduction}
%TDE, super-Eddington disk
More than a hundred tidal disruption event (TDE) candidates have been discovered through optical and X-ray surveys. TDEs offer valuable insights not only for estimating the mass and spin of massive black holes (BHs) in the centers of galaxies but also for probing accretion physics, particularly in the context of a super-Eddington accretion disk \citep{Strubbe_Optical_2009,Lodato_Multiband_2011,dai_unified_2018}. 

The newly formed TDE disk is believed to accrete at super-Eddington rates, generating strong radiation and/or magnetic pressure-driven outflows \citep{ohsuga_supercritical_2005,mckinney_three-dimensional_2014,jiang_global_2014,sadowski_global_2015,dai_unified_2018}. These outflows are thinner along the polar direction and thicker in the equatorial plane, allowing X-rays from the inner disk to escape preferentially along the polar axis. At higher angles of inclination, more X-ray emission is absorbed and the absorbed energy is reprocessed into optical/UV radiation \citep{dai_unified_2018}.

%tilted disk introduction. Many TDE disk should be tilted.
In typical TDEs, the angular momentum of the disrupted star's initial orbit is misaligned with the BH's spin, causing the resulting accretion disk to tilt relative to the BH. Under the influence of general relativity, the tilted disk undergoes Lense-Thirring (LT) precession \citep{lense_uber_1918}. For standard thin disks, the Bardeen-Petersen effect will align the inner part of the disk relatively quickly \citep{bardeen_lense-thirring_1975,papaloizou_time-dependence_1983,lodato_diffusive_2010}. However, a typical TDE disk is expected to be thick and accrete at super-Eddington rate, allowing it to precess as a rigid structure \citep{fragile_global_2007,liska_formation_2018}.

%Observational feature of precession
X-ray variability, particularly quasi-periodic oscillations (QPOs), is commonly observed in the light curves of accreting back hole systems such as BH X-ray binaries \citep[BHXRBs, ][] {fragile_bardeen-petterson_2001,miller_evidence_2005,altamirano_low-frequency_2012},  ultra-luminous X-ray sources \citep[ULXs, ][]{strohmayer_discovery_2003,mucciarelli_variable_2006} and TDEs \citep{reis_200-second_2012,saxton_long-term_2012,saxton_tidal_2012,pasham_loud_2019,pasham_lense-thirring_2024,Chen_AT2019avd_2022,wang_rapid_2024}. Some of this variability can be attributed to the LT precession in a tilted disk. For example, a jetted TDE Swift J1644 \citep{Burrows_Relativistic_2011}, and two TDE candidates, ASASSN 14li and AT 2020ocn, display QPO on their X-ray light curves, potentially indicative of precessing jet and disk structures, respectively \citep{pasham_loud_2019,pasham_lense-thirring_2024}.

%QPE
Another class of intense periodic transients, quasi-periodic eruptions (QPEs), has been identified in galactic nuclei by their very-large-amplitude and periodic X-ray bursts \citep{Miniutti_Nine_2019,Giustini_Xray_2020,Arcodia_QPE_2021,Chakraborty_Possible_2021}. Similarly to QPOs, QPEs may also arise from a precessing disk associated with an active galactic nucleus (AGN) or a TDE accreting at super-Eddington rate \citep{middleton_qpes_2025}, although alternative scenarios have been proposed. These include collisions between an orbiting body and the accretion disk surrounding a massive black hole \citep{sukova_stellar_2021, xian_x-ray_2021, franchini_qpes_2023, linial_emri_2023, tagawa_flares_2023, linial_coupled_2024, zhou_probing_2024, vurm_radiation_2024, guo_testing_2025}, disk instabilities \citep{Pan_disk_2022, kaur_magnetically_2023}, or repeating TDEs \citep{king_QPE_2022, krolik_quasiperiodic_2022, Wang_A_2022, Chen_tidal_2023}.

%J045650
Interestingly, the recently reported nuclear transient eRASSt J045650.3-203750  \citep[hereafter, J0456-20, ][]{Liu_J045650_2023,liu_rapid_2024}, identified by extended ROentgen Survey with an Imaging Telescope Array (eROSITA) in a quiescent galaxy, exhibits long-term decay with superimposed short-term quasi-periodic X-ray variability. Notably, its near-UV (NUV) and X-ray light curves display a subtle inverse relationship, a feature that is difficult to explain in the context of repeating TDEs. This behavior may instead arise from a precessing thick disk with a declining accretion rate.

%outline
In this paper, we integrate the viewing angle dependence of spectral characteristics for the super-Eddington TDE disk, as explored in \cite{dai_unified_2018} and \cite{Thomsen_Dynamical_2022}, with a precessional effect to comprehensively investigate the dynamical and radiative features of a precessing disk in TDE. Different precession patterns exhibit diverse observational signatures, which can be compared with current TDE candidates and inform future observations. To our knowledge, this is the first precessing TDE disk model that incorporates the prediction of multiwavelength light curve behavior. 

The structure of the paper is as follows: In Section \ref{sec:method}, we describe our method to calculate the spectral dependence on the viewing-angle. In Section \ref{sec:dynamical}, we explore the dynamical properties of a precessing super-Eddington disk. We then compute the light curve of a precessing TDE disk based on the combined radiative spectra and precessional effects in Section \ref{sec:Radiative}. Sections \ref{sec:discussion} and \ref{sec:conclusion} provide a discussion and summary of the results. %In Appendix, we describe the setup of our GRRMHD simulations and apply Monte Carlo radiative transfer techniques to analyze the simulation data. 

%%%%%%%%%%%%%%%%%%%%%%%%%%%%%%%%%%%%%%%%%%%%%%%%%%%%%
\section{Methodology}
\label{sec:method}
In this study, we do not simulate a physically realistic precessing super-Eddington accretion disk in TDEs. Instead, we adopt the spectral dependence on viewing angle and disk geometry for an aligned super-Eddington disk, based on simulations by \cite{dai_unified_2018} and \cite{Thomsen_Dynamical_2022}. We then impose an artificial precession by varying the viewing angle over time. However, the spectral-viewing angle relationship for a genuinely precessing, tilted disk would differ significantly. Future work should include full simulations of precessing super-Eddington TDE disks to accurately characterize their spectral properties. In this paper, we deliberately exclude the magnetically arrested disk (MAD) scenario, where the potent magneto-spin alignment mechanism has the potential to rapidly align the inner disk \citep{liska_formation_2018,chatterjee_misaligned_2025}. Our focus is solely on the disk radiative spectrum, excluding any jet components. Nevertheless, we will offer an extended discussion regarding the possible jet emission in Section \ref{subsec:jet}. We describe our method in the following sections.

\subsection{Super-Eddington accretion disk in TDE}
\label{subsec:HARMRAD}
Following the disruption of a star, the newly formed disk initially enters a super-Eddington accretion state. In this phase, intense radiation pressure and/or magnetic pressure propel the gas outward, generating a powerful outflow. This outflow is nearly axisymmetric; due to azimuthal motion, the density of the outflow is greater near the equatorial plane, where the velocity is slower, while it is lower near the polar regions, where the velocity is faster. The gas in the outflow can absorb and scatter X-ray emissions from the inner disk, subsequently re-emitting them in the optical/UV bands. This configuration results in a significant dependence of radiation on the viewing angle: an observer looking directly at the face of the disk will detect strong X-rays, whereas an observer viewing the disk edge-on will detect optical/UV emissions with weak X-rays \citep{dai_unified_2018}, as illustrated in Figure \ref{fig:Lratio_angle}.

Here, we utilize simulation results of a super-Eddington accretion disk from \cite{Thomsen_Dynamical_2022}, conducted using the general relativistic radiation magnetohydrodynamic (GR-RMHD) code HARMRAD \citep{mckinney_three-dimensional_2014}. In the simulation, the SMBH is assumed to have a mass of $10^6\ M_{\odot}$ and a dimensionless spin parameter of $a=0.8$. A quasi-steady state was achieved in the inner regions of the disk, with a mass accretion rate of $\dot{M}_{\rm acc} \simeq 12\ {\dot{M}_{\rm Edd}}$, where the Eddington mass accretion rate is defined as $\dot{M}_{\rm Edd} \equiv L_{\rm Edd}/(\eta_{\rm NT} c^2)$, with $L_{\rm Edd}$ being the Eddington luminosity and $\eta_{\rm NT} \simeq 0.12$ the nominal accretion efficiency for a Novikov-Thorne thin-disk solution \citep{novikov_astrophysics_1973}.

\subsection{Spectra in different inclinations}
\label{subsec:sedona}
To determine the escaped luminosity at different wavelengths, we utilize the result in \cite{Thomsen_Dynamical_2022}, obtained by the Monte Carlo radiative transfer code SEDONA \citep{kasen_time-dependent_2006} to perform radiation transport calculations. 

%The radiation transport is calculated using Monte Carlo method which includes scattering, bound-bound, bound-free, and free-free radiative processes. Additionally, the calculations include non-local thermal equilibrium opacity \citep{roth_x-ray_2016} and electron Comptonization effects \citep{roth_what_2018}. The gas is assumed to consist of only H, He, and O with solar abundances. Source photons are injected from the inner boundary with a blackbody luminosity of $10^{44}\ {\rm erg/s}$ and a temperature of $10^6\ {\rm K}$. 

Following the method in \cite{Thomsen_Dynamical_2022}, we explore the spectra for a range of inclinations, from $15^{\degree}$ to $70^{\degree}$. For each inclination, the gas density, temperature, and velocity are averaged within $\pm 5^{\degree}$ of the inclination from the GR-RMHD simulation, to generate a spherical input profile for SEDONA. The results are shown in Figure \ref{fig:spectrum}, which highlights the $0.2 - 10$ keV X-ray band and the blackbody fitting in the optical band.

%Additionally, the jet density is reduced by a factor of 100 before averaging, and the velocity of the disk inflow region is set to zero \cite[see the detailed setup in ][]{Thomsen_Dynamical_2022}.

%The gas temperatures, ionization states, bound electron states, and radiative transfer solutions are iteratively calculated until the output spectra converge, typically requiring 20-30 iterations. 

\begin{figure*}		
\centering
 \begin{minipage}[t]{0.45\textwidth}
\centering
 \includegraphics[scale=0.35]{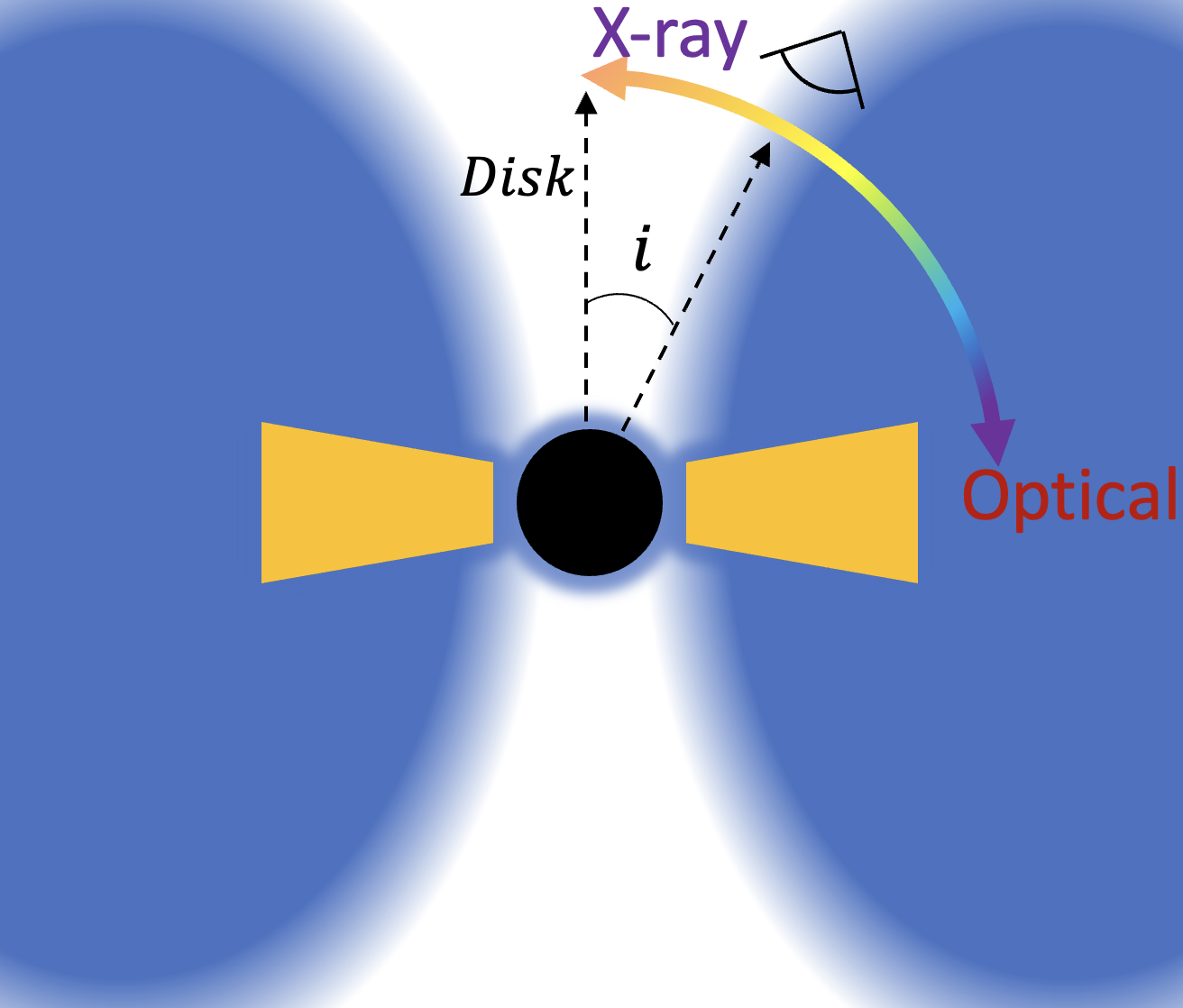}
 \end{minipage}
\begin{minipage}[t]{0.45\textwidth}
\centering
 \includegraphics[scale=0.5]{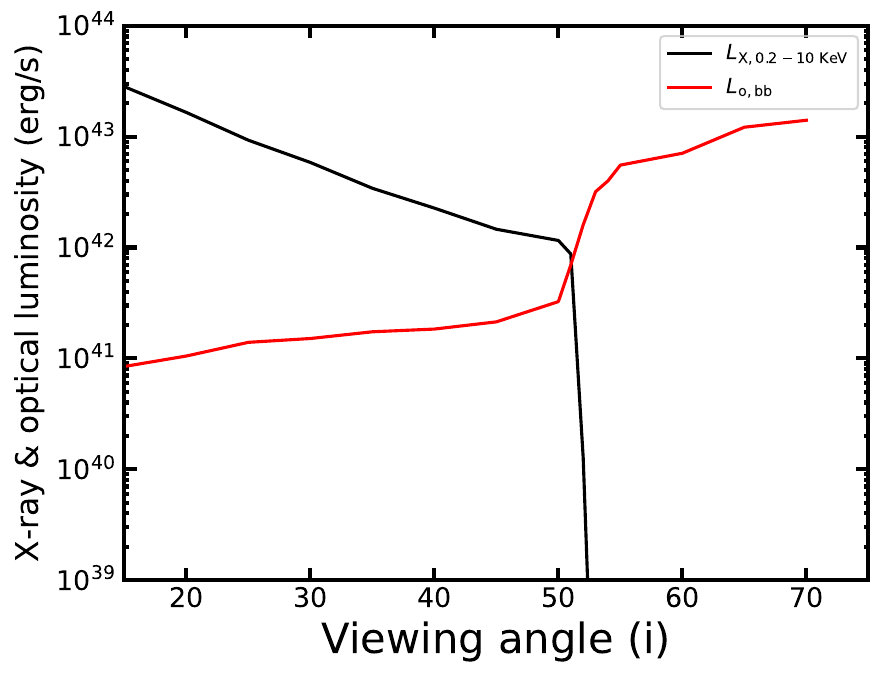}
 \end{minipage}
    \caption{Left: Schematic of a super-Eddington accretion disk with outflowing material (blue shaded region). When the observer's line of sight aligns with the outflow-cleared funnel, bright X-ray emission is detected. Conversely, observations along the equatorial plane are dominated by optical emission due to reprocessing. Right: $L_{\rm X, 0.2 - 10\ {\rm keV}}$ and optical luminosity $L_{\rm o, bb}$ versus viewing angle $i$.}
    \label{fig:Lratio_angle}
\end{figure*}

Figure \ref{fig:Lratio_angle} illustrates the X-ray luminosity in the $0.2 - 10$ keV band ($L_{\rm X, 0.2 - 10\ {\rm keV}}$) and the optical luminosity ($L_{\rm o, bb}$), derived from blackbody fitting the \textit{Swift}/UVOT band (v, b, u, UVW1, UVM2, and UVW2 filters), as a function of the viewing angle. In particular, as the viewing angle exceeds $\sim 50^{\degree}$, the X-ray luminosity undergoes a sharp decline, while $L_{\rm o, bb}$ exhibits a slight increase. X-ray luminosity evolves more rapidly than optical emission, while optical luminosity exhibits minimal changes within the $0^{\degree}-50^{\degree}$ and $55^{\degree}-70^{\degree}$ ranges.

%The corresponding blackbody temperature remains relative constant across various viewing angles with changes less than a factor of two. This consistency aligns with observational findings for optical TDE candidates, which exhibit constant temperatures as luminosity evolves.

In the following sections, we proceed under the assumption of a rigidly precessing disk exhibiting the aforementioned viewing angle dependency to calculate the corresponding radiative features.

\section{Viewing-angle modulation of a Precessing Super-Eddington Disk} 
\label{sec:dynamical}
The disruption of a star by an SMBH typically forms a super-Eddington accretion disk with powerful disk winds \citep{Strubbe_Optical_2009,Lodato_Multiband_2011,dai_unified_2018}. As demonstrated by \cite{dai_unified_2018}, the viewing angle plays a crucial role in observational signatures: near-polar orientations allow direct detection of inner-disk X-ray emission, while near-edge-on views lead to X-ray absorption by the thick outflow, with reprocessed emission dominating the optical/UV bands \cite[see also ][]{qiao_early_2025}.

The nascent accretion disk is typically misaligned with the SMBH's spin axis because of the random orientation of the stellar orbit. This misalignment induces LT precession via general relativistic effects \citep{lense_uber_1918}. Simulations suggest that thick, super-Eddington disks can precess coherently as rigid bodies \citep{fragile_global_2007,liska_formation_2018}.

\begin{figure*}		
\centering
 \begin{minipage}[t]{0.45\textwidth}
\centering
 \includegraphics[scale=0.17]{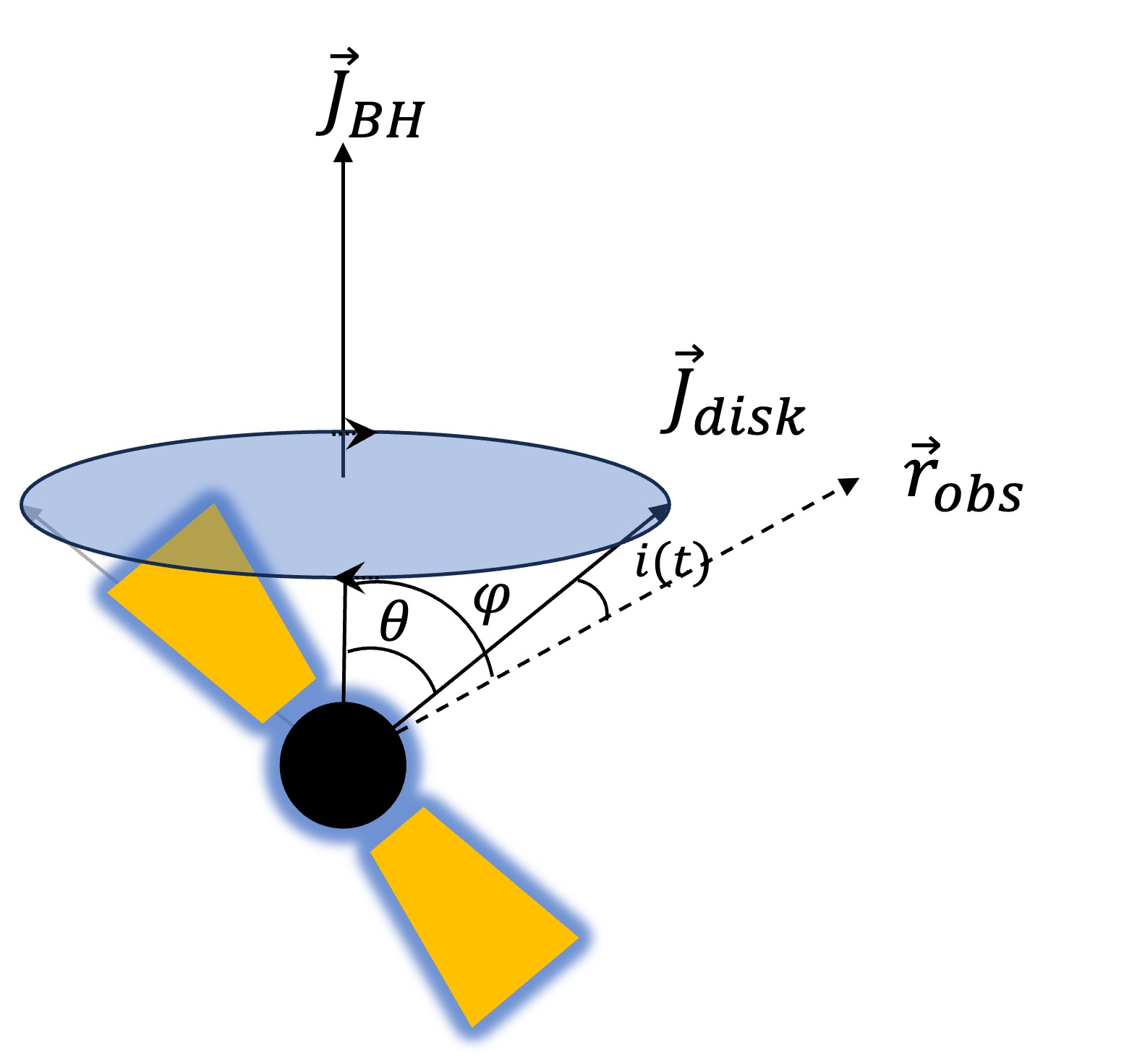}
 \end{minipage}
 \hspace*{1.0cm}
\begin{minipage}[t]{0.45\textwidth}
\centering
 \includegraphics[scale=0.26]{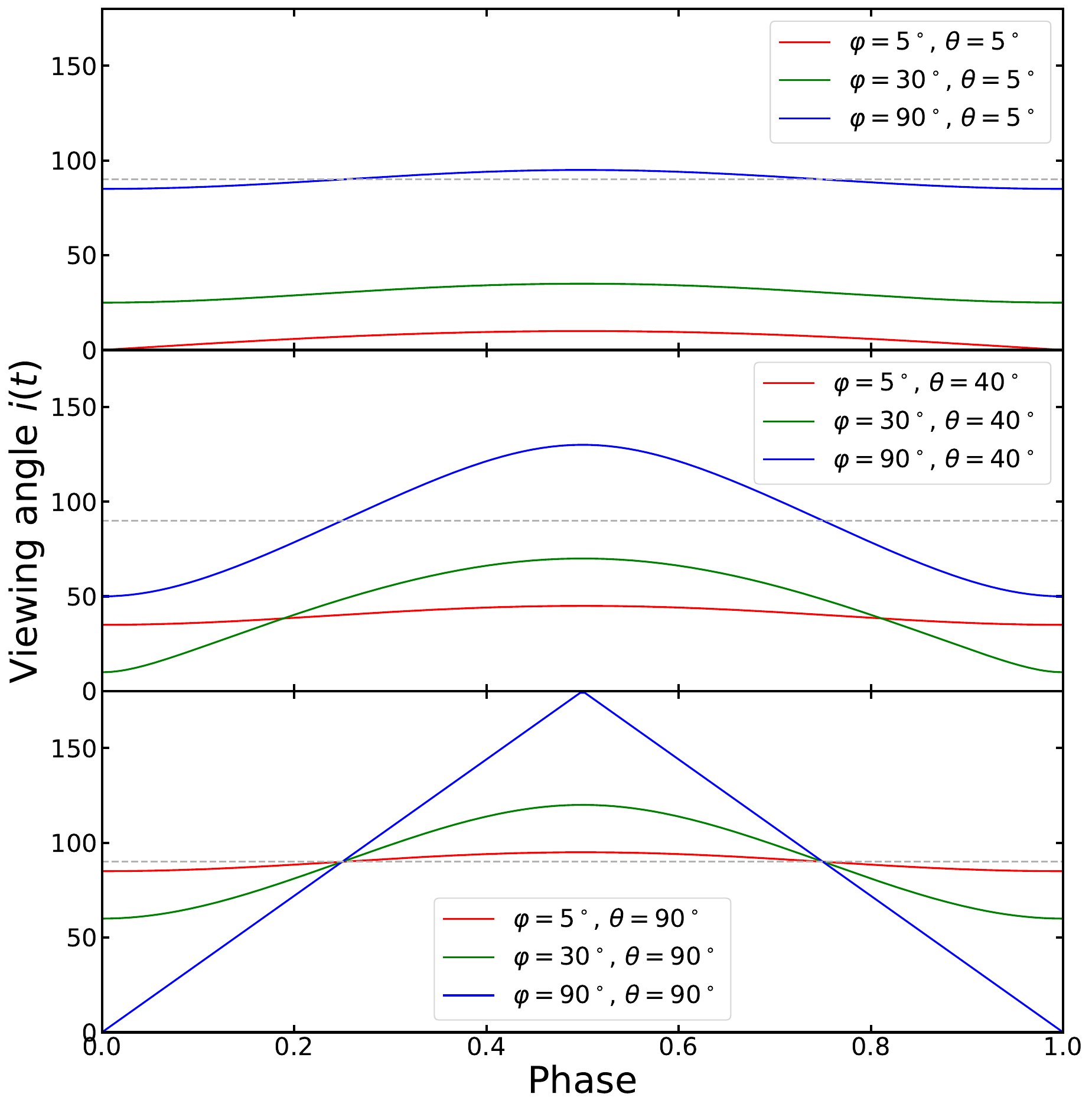}
 \end{minipage}
    \caption{Left: Sketch of disk precession geometry. As the disk precesses around the BH's spin vector ($\vec{J}_{\rm BH}$), the viewing angle $i(t)$ between the observer's line of sight ($\vec{r}_{\rm obs}$) and the disk orientation ($\vec{J}_{\rm disk}$) varies. The angle $\theta$ denotes the inclination between $\vec{J}_{\rm BH}$ and $\vec{J}_{\rm disk}$, while $\varphi$ represents the angle between $\vec{J}_{\rm BH}$ and $\vec{r}_{\rm obs}$. Both $\theta$ and $\varphi$ remain nearly constant during the procession cycle. Right: Evolution of viewing angle $i(t)$ of a precessing disk over one precession period. Upper, middle and lower panels illustrate the small, medium and large $\theta$ cases, respectively. The dashed gray line indicates the $90^{\degree}$ viewing angle, beyond which the opposite side of the disk becomes visible as the disk viewing angle crosses this value. The blue line in the lower panel corresponds to the most extreme case with $\theta = \varphi = 90^{\degree}$.}
    \label{fig:viewing}
\end{figure*}

The observed flux modulation amplitude from such a precessing disk depends on the viewing angle between the observer's line of sight and the disk, as well as the spin orientation of the SMBH, as illustrated in Figure \ref{fig:viewing}. 

As the disk precesses, the observed flux evolves with the angle $\theta$ between the line of sight and the disk orientation. From geometric considerations,  $i(t)$ evolves as
\begin{equation} \label{eq:cos_i}
    \cos{[i(t)]} = \cos{\varphi} \cos{\theta} + \sin{\varphi} \sin{\theta} \cos{(\omega t)},
\end{equation}
where $\varphi$ is the angle between the observer's line of sight and BH's spin axis, $\theta$ is the disk tilt angle relative to the spin axis, and $\omega$ is the precession angular velocity. In a precession cycle, $i(t)$ ranges from 
\begin{equation} \label{eq:irange}
    i_{\rm min}=|\varphi - \theta|\ {\rm to}\ i_{\rm max} = \min(\varphi + \theta, 2\pi-\varphi - \theta),
\end{equation}
where $\theta \in [0, 180^{\degree}]$ and $\varphi \in [0, 180^{\degree}]$.

The observed variability strongly depends on the geometry of the system. For a small tilt angle ($\theta \sim 0^{\degree}$), the disk remains nearly aligned with the black hole's spin axis, resulting in minimal precession. When the observer's line of sight is nearly aligned with the spin axis ($\varphi \sim 0^{\degree}$), the viewing angle remains approximately constant. In both cases, the flux variations induced by disk precession are negligible. 

When both $\varphi$ and $\theta$ are large enough, $i(t)$ can exceed $90^{\degree}$ during precession. This causes the opposite side of the disk to become visible, leading to more pronounced variability.

Figure \ref{fig:viewing} demonstrates the evolution of the viewing angle for different configurations $\theta$ and $\varphi$. For the small $\theta$ case, the viewing angle variation remains small as the disk maintains approximate alignment with the SMBH's spin axis. In the extreme case where the viewing angle crosses $90^{\degree}$,  observers see both faces of the disk during each precession cycle, resulting in two edge-on orientations per period.

%%%%%%%%%%%%%%%%%%%%%%%%%%%%%%%%%%%%%%%%%
\section{Radiative Features}
\label{sec:Radiative}
\subsection{X-ray and optical luminosity modulation in a precession cycle}
\label{subsec:X-ray_optical}
The observed emission properties are critically dependent on both the disk structure and the viewing geometry. For a standard thin disk, the flux follows a simple denpendence of $\propto \cos{\theta}$ with minimal spectral changes. However, super-Eddington accretion disks exhibit more complex behavior as a result of their powerful outflows. When viewed pole-on, observers detect strong X-ray emission from the inner disk and potentially jets. At higher inclinations, the outflow obscures X-rays, leading to optical/UV-dominated emission through reprocessing.

\begin{figure*}		
\centering
\begin{minipage}[t]{0.8\textwidth}
\centering
 \includegraphics[scale=0.1]{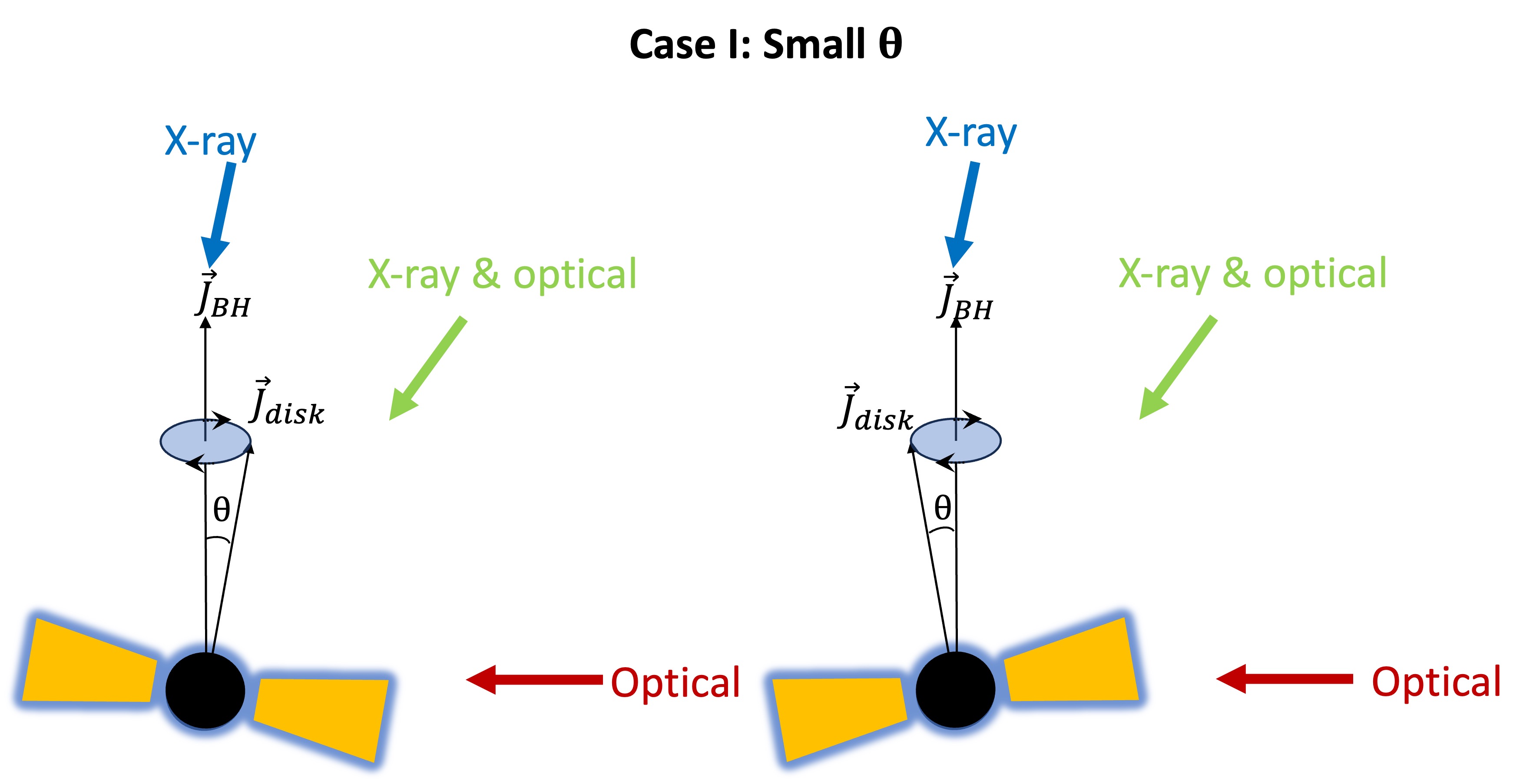}
 \end{minipage}
 \begin{minipage}[t]{0.8\textwidth}
\centering
 \includegraphics[scale=0.1]{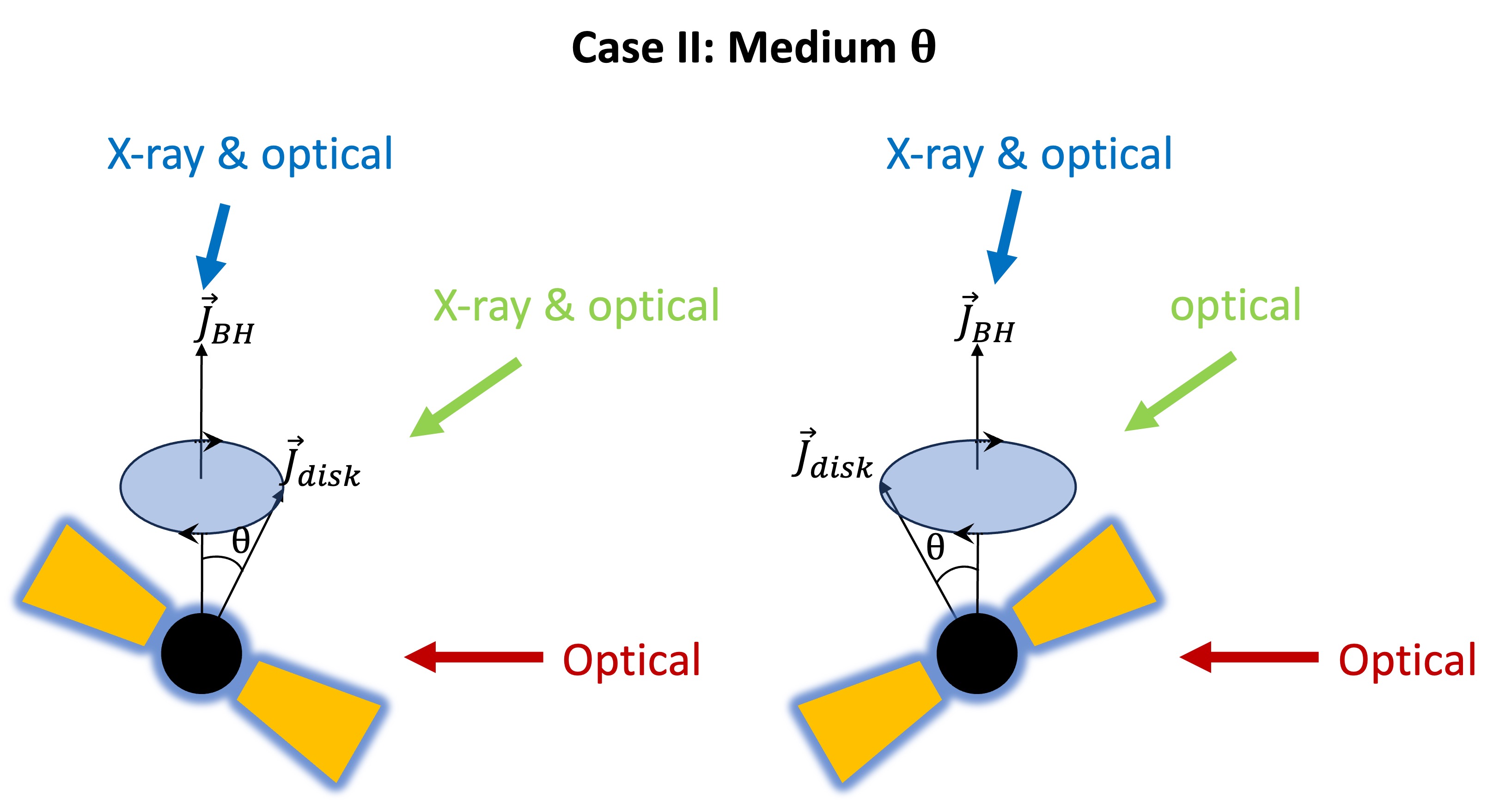}
 \end{minipage}
 \begin{minipage}[t]{0.8\textwidth}
\centering
 \includegraphics[scale=0.1]{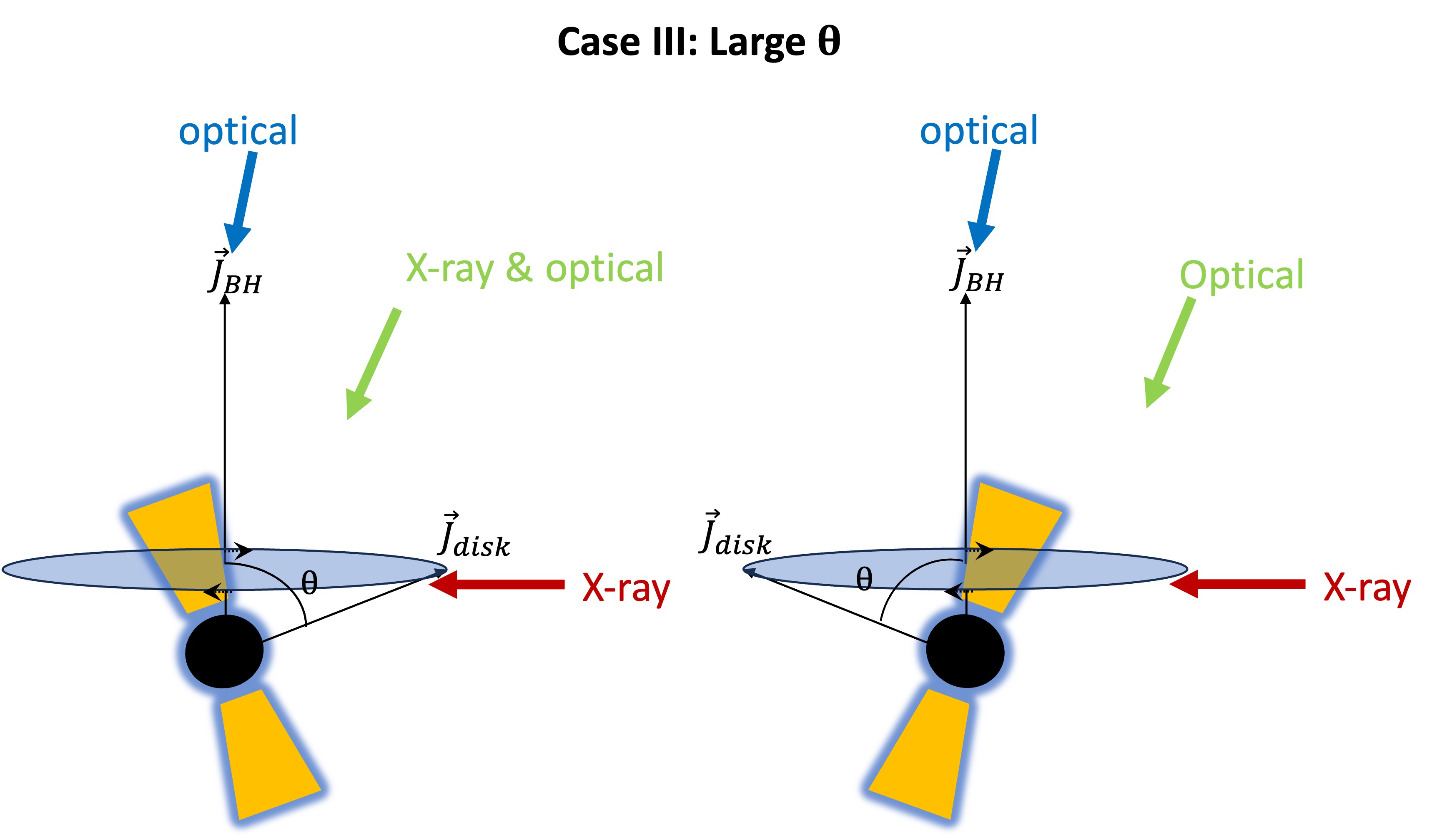}
 \end{minipage}
    \caption{Upper, middle and lower panels depict three distinct cases corresponding to different tilted angle $\theta$. The left and right panels illustrate the disk's positions at the two antipodal points of the precession cycle. The direction of precession is indicated by a circular arrow. The colored and labeled arrows elucidate the radiation patterns detectable by an observer as they scan different directions within a precession cycle.}
    \label{fig:viewing_cases}
\end{figure*}

We identify three distinct precession modes based on $\varphi$ and $\theta$ configurations, as illustrated in Figure \ref{fig:viewing_cases}

\textbf{Small tilt angle}:
Spin-aligned observer ($\varphi \simeq 0^{\degree}$): X-ray-dominated emission with minimal variability (e.g., some TDEs with steady X-ray emission). Intermediate $\varphi$: Simultaneous X-ray and optical emission with low variability  \citep[e.g., ASASSN-14li][]{holoien_six_2016,brown_long_2017} and ASASSN-20qc \citep{pasham_case_2024} Notably, these sources show quasi-periodic X-ray variability that potentially indicates precession. Edge-on observer ($\varphi \simeq 90^{\degree}$): Optical-dominated and X-ray-weak emission without variability \citep[e.g., ASASSN-14ae ][]{Holoien_ASASSN_2014}.
        
\textbf{Moderate tilt angle}: Spin-aligned observer: Similar to small-$\theta$ case with mixed X-ray/optical emission in the medium inclination. Intermediate $\varphi$: exhibit periodic transitions between X-ray-bright and optical-dominated states during the precession cycle \citep[e.g., J0456-20 ][]{liu_rapid_2024}. Edge-on observer: Optical-dominated emission with variability \citep[e.g., ASASSN-14ko ][]{Payne_14ko_2021,Payne_14ko_2022,Liu_Tidal_2023}. 

\textbf{Large tilt angle}: Spin-aligned observer: Optical-dominated with minimal variability, similar to the small $\theta$ case in the vertical direction. Intermediate $\varphi$: optical-dominated TDE with a large modulation amplitude in optical, similar to the medium $\theta$ case in medium inclination. Edge-on observer:  Both X-ray and optical emissions undergo extreme flux modulation. As the accretion disk transitions between face-on and edge-on orientations, the emission alternates between being X-ray-dominated and optical-dominated, respectively.

Subsequently, we employ these results to model the observed light curve from a precessing disk, where the viewing angle varies periodically. The viewing angle may vary in the polar direction ($\sim 0^{\degree}$) or in the equatorial direction ($\sim 90^{\degree}$); however, it is not feasible to acquire spectra directly at these extreme viewing angles. At high inclinations, approaching an edge-on orientation, the optical depth becomes excessively large, rendering post-processing with SEDONA impractical. In contrast, at low inclinations, the results are compromised by contamination from the jet component, thus undermining their reliability. Instead, we must make assumptions to extrapolate the luminosity-angle relationship to these ranges. Here, we assume that the luminosity within $0^{\degree}-15^{\degree}$ is equal to the value at $15^{\degree}$, and the luminosity within $70^{\degree}-90^{\degree}$ is equal to the value at $70^{\degree}$.

\begin{figure*}		
\centering
 \includegraphics[scale=0.6]{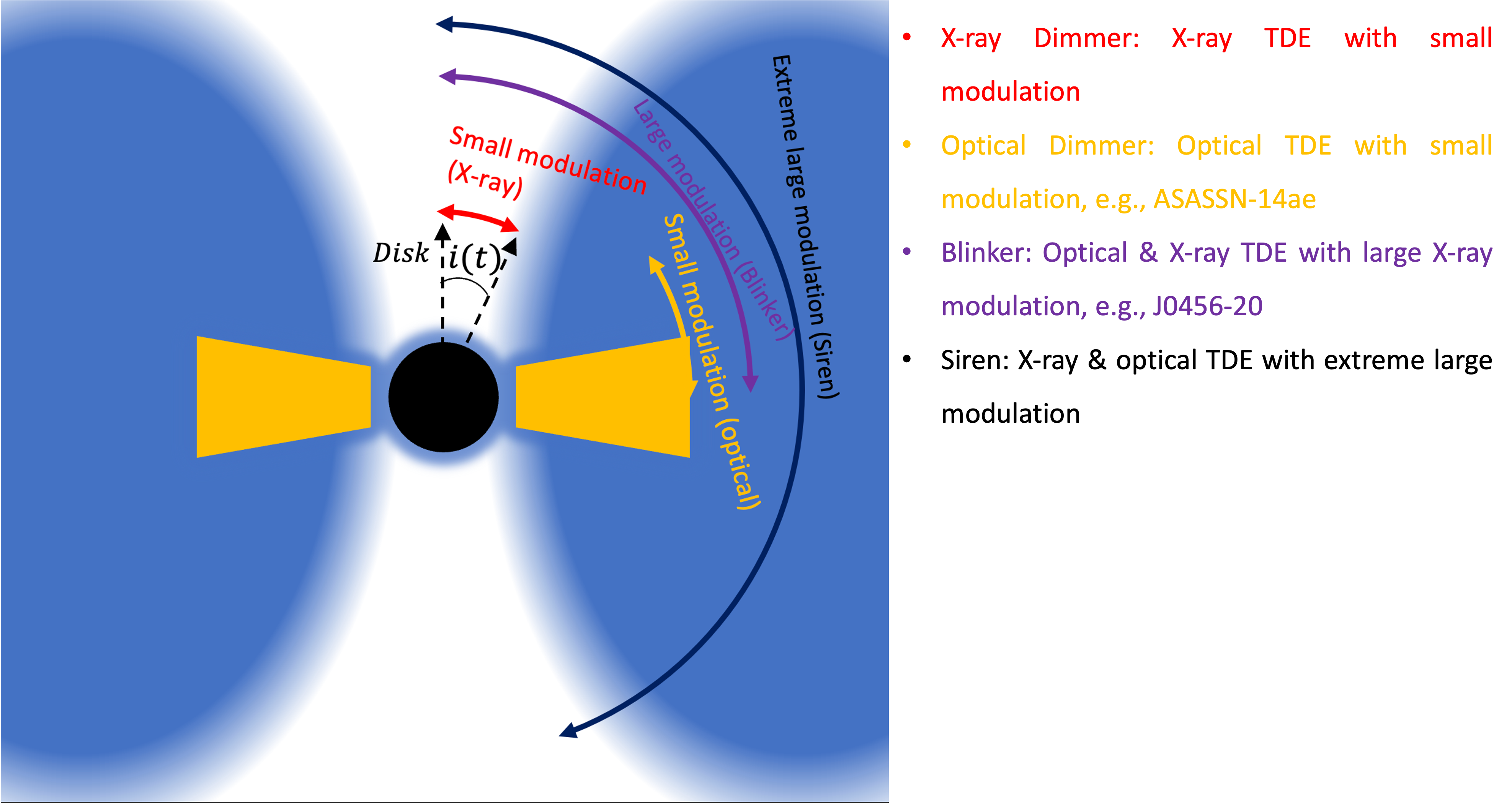}
    \caption{Schematic of observed precession patterns in an super-Eddington accretion disk. As the viewing angle varies across different ranges, the observer detects distinct signatures of disk precession, as indicated by the double arrows. A summary of these patterns is provided in Table \ref{tab:patterns}.}
    \label{fig:pattern}
\end{figure*}

\begin{figure*}		
\centering
\begin{minipage}[t]{0.45\textwidth}
\centering
 \includegraphics[scale=0.25]{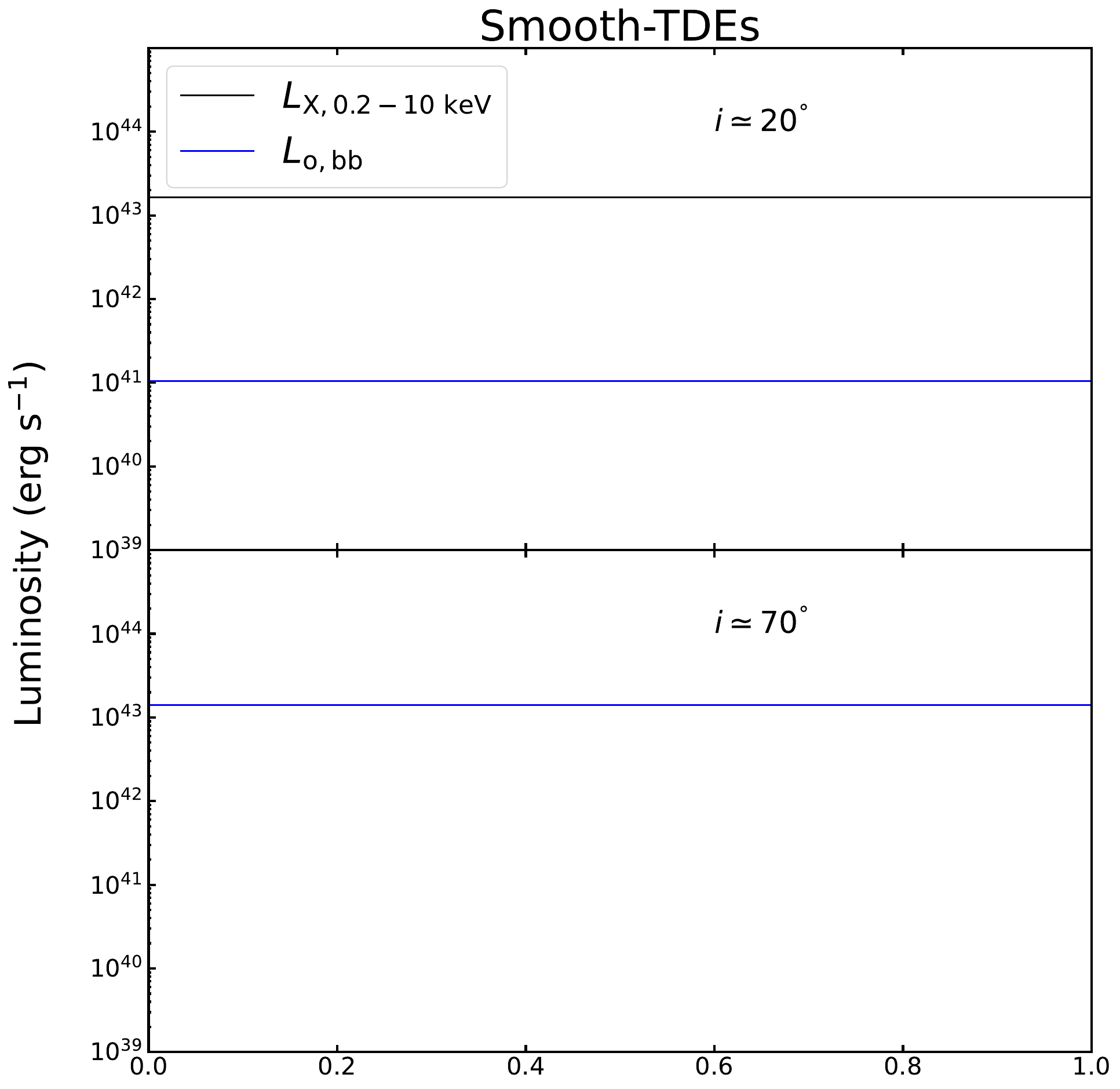}
 \end{minipage}
\begin{minipage}[t]{0.45\textwidth}
\centering
 \includegraphics[scale=0.25]{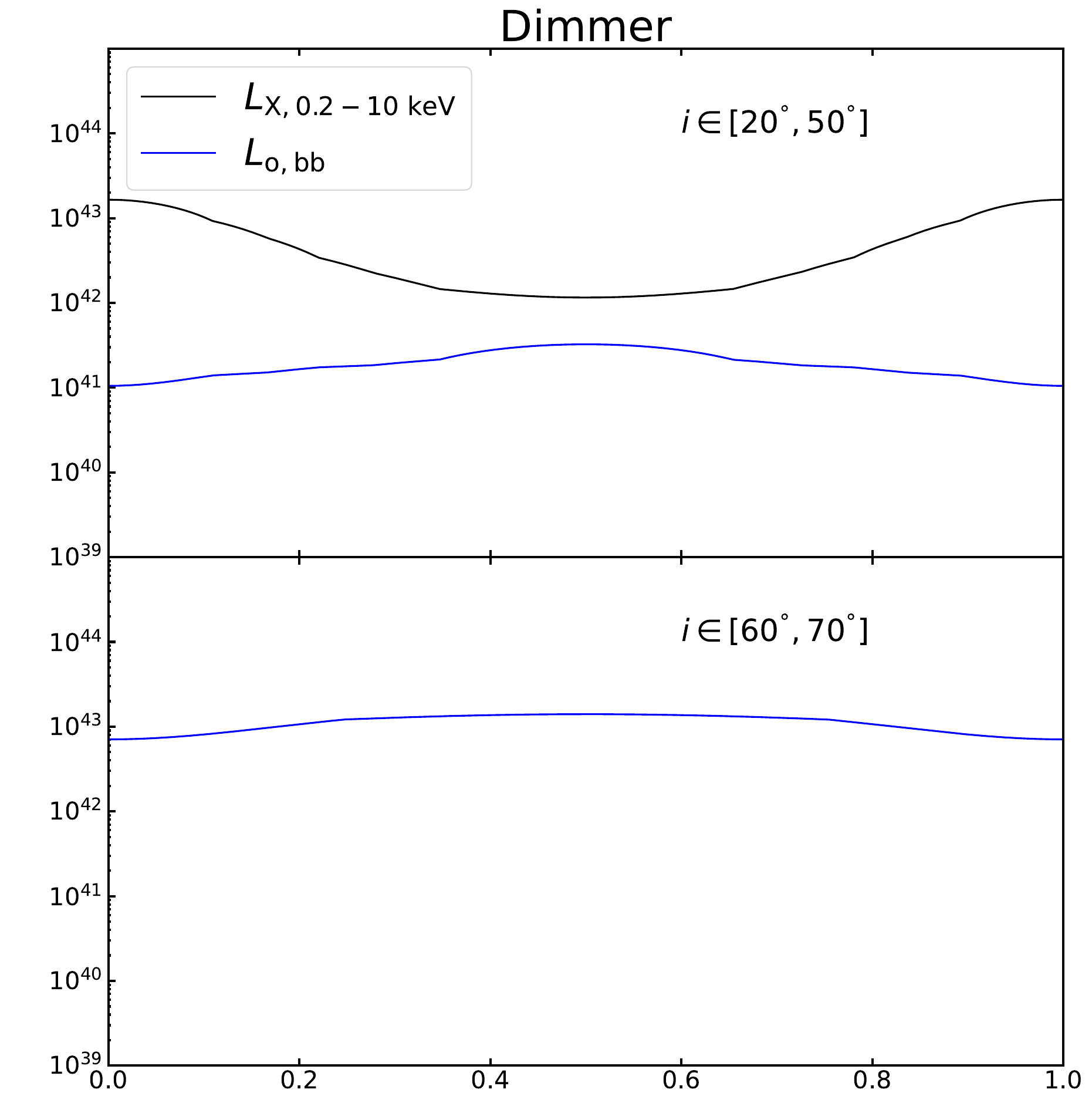}
 \end{minipage}
\begin{minipage}[t]{0.45\textwidth}
\centering
 \includegraphics[scale=0.25]{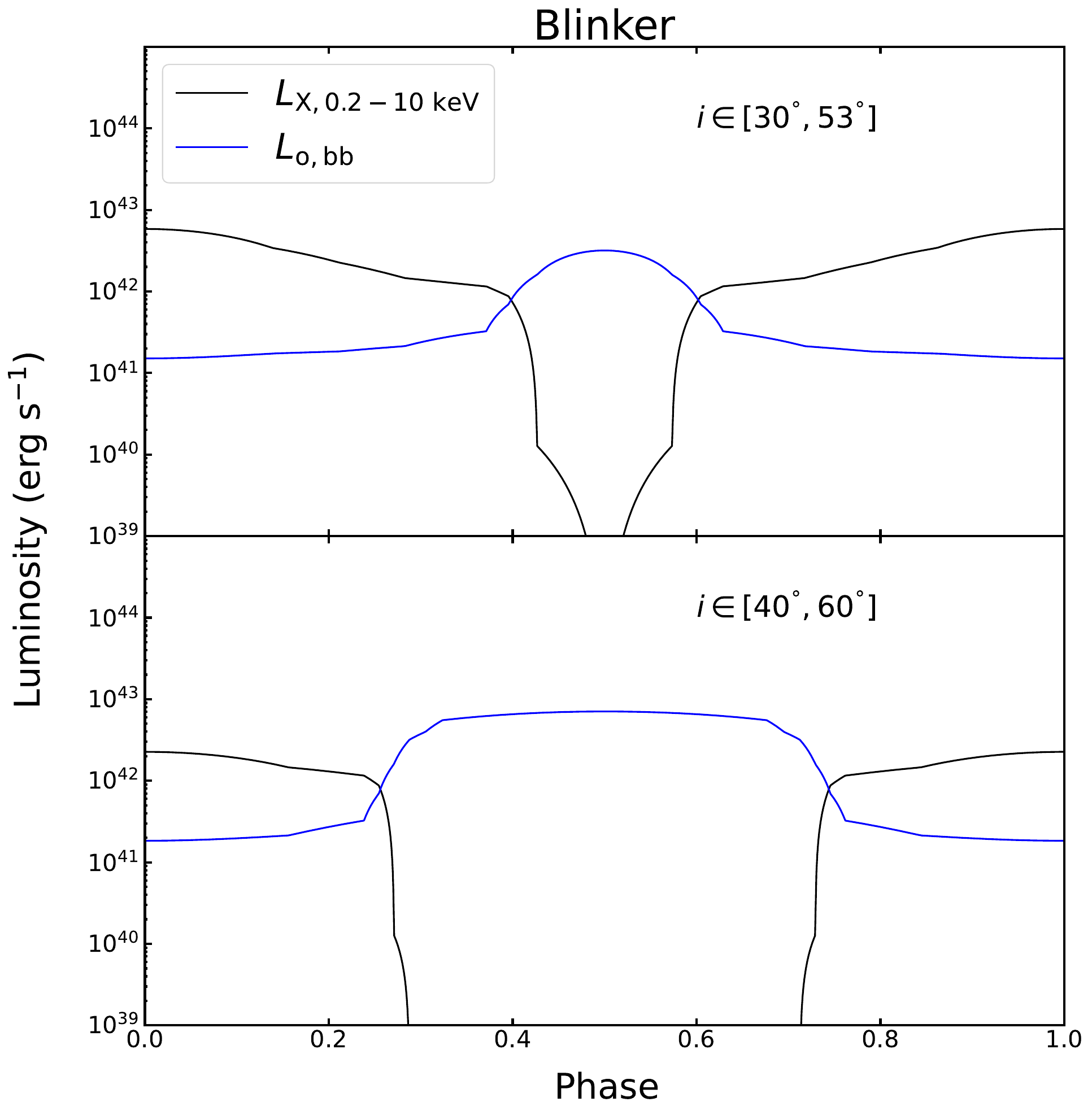}
 \end{minipage}
\begin{minipage}[t]{0.45\textwidth}
\centering
 \includegraphics[scale=0.25]{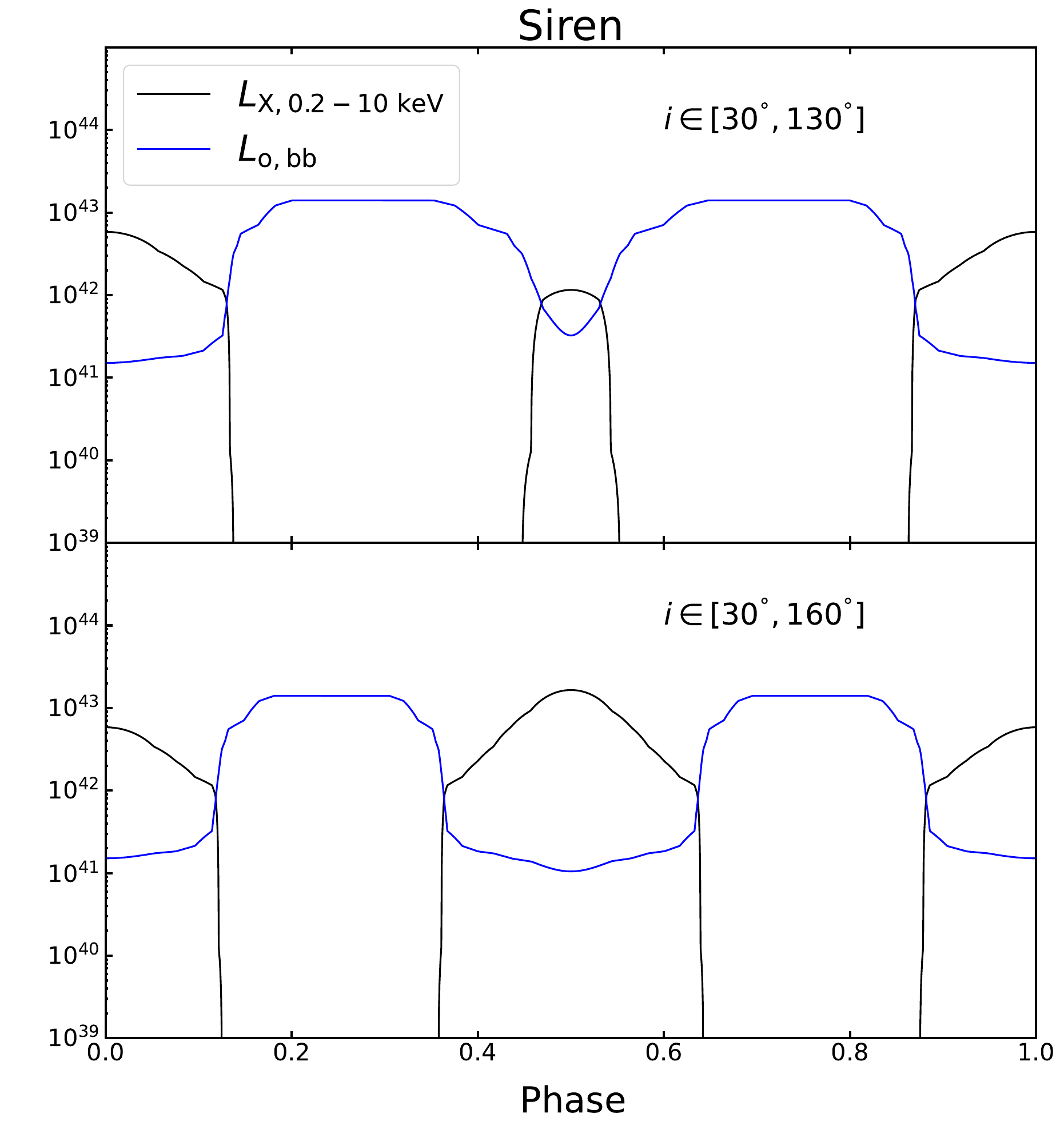}
 \end{minipage}
    \caption{The evolutions of X-ray luminosity $L_{\rm X, 0.2 - 2\ {\rm keV}}$ (black lines) and optical luminosity $L_{\rm o, bb}$ (blue lines) due to disk precession over one precession period for different patterns. Each panel shows the results corresponding to distinct viewing angle ranges. Characteristic examples of the smooth-TDEs, the Dimmer, the Blinker, and the Siren are illustrated in the top-left, top-right, bottom-left, and bottom-right panels, respectively.}
    \label{fig:Lx_Lop_phase}
\end{figure*}

In Figure \ref{fig:viewing_cases}, we see that the precession mode depends on both $\theta$ and $\varphi$, with some modes exhibiting similar radiative characteristics. The luminosity evolution is directly dependent on the viewing angle range during a precession cycle. When the viewing angle range is small, i.e., $i_{\rm max}/i_{\rm min} \lesssim 2$, and does not cross the rapid transition point at $50^{\degree}$, the X-ray luminosity varies by less than one order of magnitude, resulting in weak modulation. Otherwise, the modulation is more pronounced. Based on these radiative characteristics, we classify the precession modes into distinct patterns according to the range of viewing-angle variations, as depicted in Figure \ref{fig:pattern}. The corresponding luminosities for each pattern are illustrated in Figure \ref{fig:Lx_Lop_phase}.

We can distinguish 4 different cases. First, for cases with a constant viewing angle, the emission would not show precession-induced variability. Second, for cases where the viewing angle undergoes only minor changes, the dominant emission is X-ray if the viewing angle is near the polar region; otherwise, it is optical. These cases correspond to X-ray or optical TDEs with small variability. Third, when the viewing angle shifts between the thick outflow region and the polar region, we observe a significant modulation amplitude in the X-ray light curve. Finally, If the viewing angle range is even larger and spans $90^{\degree}$, the disk edge can be observed twice within a single cycle, resulting in two optical peaks during one phase. We designate these 4 cases as \textit{smooth-TDE}, \textit{Dimmer}, \textit{Blinker}, \textit{Siren}, respectively.

%\begin{figure*}		
%\centering
% \includegraphics[scale=0.5]{Lx_Lop_phase.pdf}
%    \caption{Characteristic examples of the evolutions of X-ray luminosity $L_{\rm X, 0.2 - 2\ {\rm keV}}$ (black lines) and optical luminosity $L_{\rm o, bb}$ (blue lines) due to disk precession over one precession period. Each panel shows the results corresponding to distinct viewing angle ranges. The left and right panels present the no/weak and large/extreme large modulation patterns, respectively. A narrow range of viewing angles arises when observing a precessing disk due to Small $\theta$ or small $\varphi$, resulting in the no/weak modulation. Otherwise, the wide range of viewing angles leads to large/extreme large amplitude of modulation.}
%    \label{fig:Lx_Lop_phase}
%\end{figure*}

These findings elucidate the radiative evolution of a precessing disk in a stable stage. However, in the context of TDE, the accretion rate undergoes a rapid increase followed by a gradual decline. Consequently, the light curve should reflect the interplay between these two effects. In the forthcoming section, we will delve into the analysis of the light curve emanating from a precessing disk in a declining accretion rate.

\begin{table*}
\centering
\caption{Summary of the precession patterns of TDEs}
\label{tab:patterns}
\begin{tabular}{lccc}
\toprule
Name & Type & Viewing angle & Num. of modulation \\
\midrule
Smooth-TDEs & No modulation & $i_{\rm min} \sim i_{\rm max}$ & 0 \\
Dimmer & Small modulation in X-ray & $i_{\rm min} < i_{\rm max} \lesssim 50^{\degree}$ or $130^{\degree} \lesssim i_{\rm min} < i_{\rm max}$ & 1\\
 & Small modulation in optical & $50^{\degree} \lesssim i_{\rm min} < i_{\rm max} \lesssim 130^{\degree}$ & 1 \\
Blinker & Large modulation & $i_{\rm min} \lesssim 50^{\degree}$ and $50^{\degree} \lesssim i_{\rm max} \lesssim 130^{\degree}$& 1 \\
Siren & Extreme large modulation & $i_{\rm min} \lesssim 50^{\degree}$ and $i_{\rm max} \gtrsim 130^{\degree}$ & 2 \\
\bottomrule
\end{tabular}
\tablecomments{\footnotesize We classify the four precession patterns as listed in Column 1. Columns 2-4 specify the modulation type, the corresponding viewing angle range, and the number of modulation peaks per precession cycle, respectively. Figure \ref{fig:pattern} illustrate these four patterns.}
\end{table*}

\subsection{Light curve with declining accretion rate}
\label{subsec:LC_Macc}
The evolution of multiwavelength emission, particularly X-ray and optical emission, in observed TDE candidates is complex, as these emissions do not always follow the same trend. For simplicity, we assume that their overall behavior adheres to a $-5/3$ power-law decay, consistent with the typical mass fallback rate observed in TDEs. We express the X-ray luminosity and optical luminosity with declining accretion rate as follows:
\begin{equation}
    \label{eq:L_x}
    \tilde{L}_{\rm X,0.2-2\ keV}(t) \simeq L_0 \left(\frac{t+t_{\rm fb}-t_0}{t_{\rm fb}}\right)^{-5/3} \times L_{\rm X,0.2-2\ keV}
\end{equation}
\begin{equation}
    \label{eq:L_op}
    \tilde{L}_{\rm o, bb}(t) \simeq L_0 \left(\frac{t+t_{\rm fb}-t_0}{t_{\rm fb}}\right)^{-5/3} \times L_{\rm o, bb}.
\end{equation}
Here, constant $L_0$ is the re-scale factor for luminosity, while $t_0$, $t_{\rm fb}$ denote the zero-point time and the timescale of mass fallback, respectively. $t_{\rm fb}$ depends on the black hole mass, pericenter radius, stellar mass and stellar structure, and can be estimated by
\begin{equation}
    t_{\rm fb} \simeq 41\ M_6^{1/2} r_*^{3/2} m_*^{-1}\ {\rm day},
\end{equation}
where $M_{\rm h} = M_6 \times 10^6 M_{\odot}$, $M_* = m_* \times M_{\odot}$ and $R_* = r_* \times R_{\odot}$ represent the black hole mass, stellar mass, and radius, respectively.

Note that for the sake of simplicity, we calculate the light curve assuming a declining accretion rate by proportionally reducing the luminosity while preserving the spectral shape. In reality, super-Eddington outflows can obscure a greater proportion of X-rays in higher accretion rate \citep{dai_unified_2018}. However, we choose not to incorporate this effect in the current study, reserving its exploration for future research endeavors.

Equations (\ref{eq:L_x}) -- (\ref{eq:L_op}) suggest that the overall light curve exhibits a power-law decay, whereas the multiwavelength emissions undergo periodic modulation due to precession. 

\begin{figure*}		
\centering
 \includegraphics[scale=0.5]{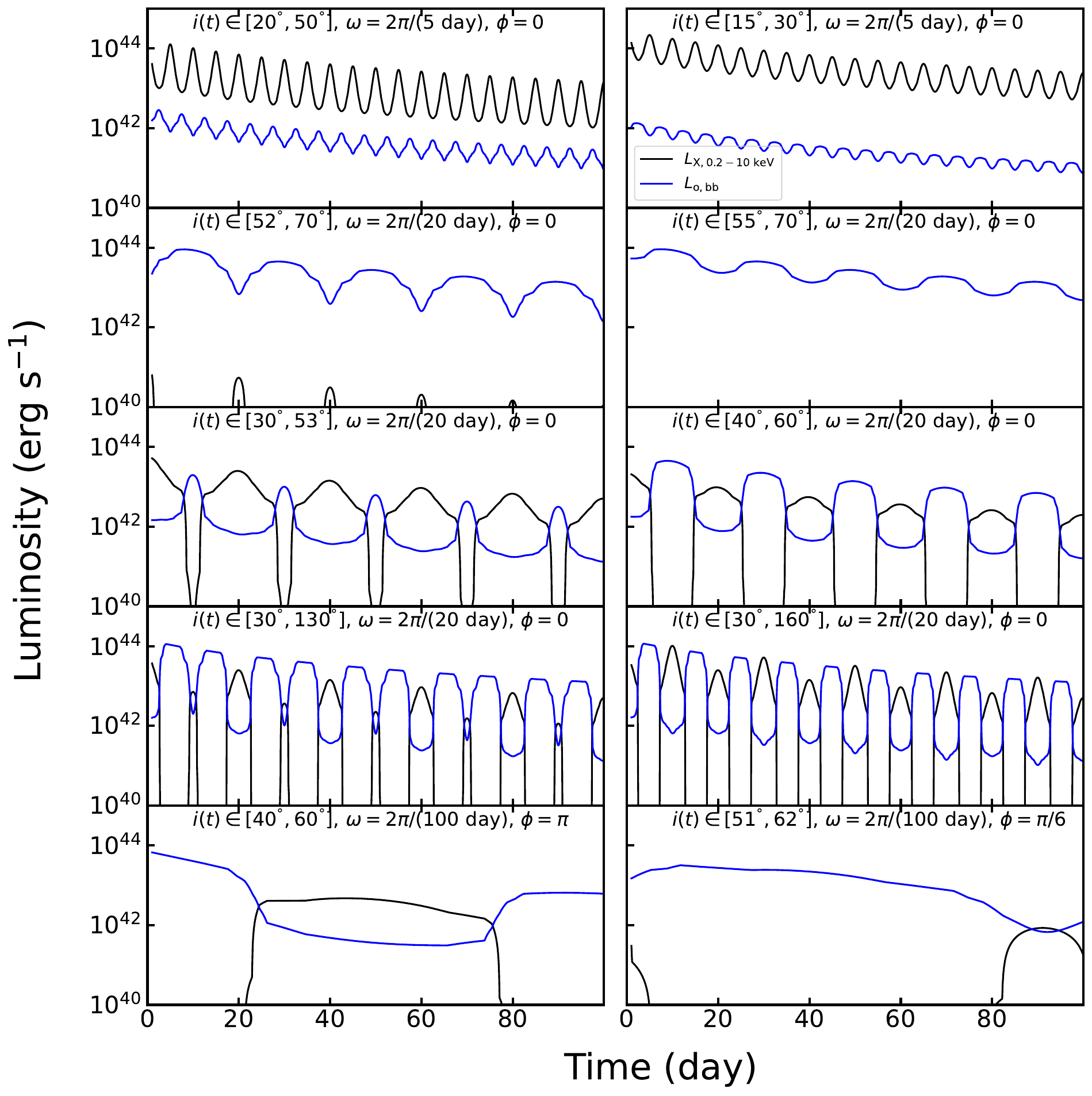}
    \caption{Charateristic examples of evolutions of X-ray luminosity $\tilde{L}_{\rm X, 0.2 - 2\ {\rm keV}}$ (black lines) and optical luminosity $\tilde{L}_{\rm o, bb}$ (blue lines) are represented, combining the effect of disk precession and declining accretion rate. Each panel shows the results obtained with distinct parameters, as labeled in the figure. Panels in rows 1-4 correspond to rapid precession, while row 5 illustrates the slow precession case. A detailed discussion on these findings can be found in Section \ref{subsec:LC_Macc}.}
    \label{fig:Lx_Lop_dec}
\end{figure*}

In Figure \ref{fig:Lx_Lop_dec}, we present five different representative
cases with varying parameters in different rows. All cases share the same $t_0=0$, $t_{\rm fb} =30\ {\rm day}$ and $L_0 = 10$, but differ in their precession parameters, as indicated in each panel. In the following, we describe the characteristics of each case, presented in order from top to bottom.
\begin{enumerate}
    \item[1.] This case illustrates a TDE with rapid and minor variability in the X-ray light curve, like the QPO signal, while the optical emission shows little variability and follows the X-ray decay. We designate this case as the ``\textit{Dimmer}''. This behavior resembles the TDE candidate ASASSN-14li \citep{holoien_six_2016, brown_long_2017}, which exhibits both X-ray and optical emissions, with X-rays displaying a QPO signal attributed to a precessing disk \citep{pasham_loud_2019}. Another candidate, ASASSN-20qc, also shows minor X-ray variability, possibly due to quasi-periodic outflows from a binary black hole TDE modulating the X-ray emission \citep{pasham_case_2024}.

    \item[2.] In this case, the viewing direction is close to the disk edge, resulting in optical-dominated emission with small variability. We also designate this case as the ``\textit{Dimmer}''.  A similar source is ASASSN-14ko, which displays periodic optical flares \citep{Payne_14ko_2021,Payne_14ko_2022}. However, the overall luminosity of ASASSN-14ko does not show a discernible decreasing trend, suggesting that recurring flares are more likely to be attributable to repeated TDE rather than to a single TDE with modulation \citep{Liu_Tidal_2023,huang_dissonance_2023}. Interestingly, the host galaxy of ASAASN-14ko is classified as an AGN \citep{holoien_asas-sn_2014}. Therefore, we consider that the recurring flares of ASASSN-14ko may arise from the precession of the AGN disk, but the rigid-body precession period could be too long for detection. Alternatively, the AGN disk may not be undergoing rigid-body precession due to its large size, instead, a warp wave-like disturbance could be propagating within the disk, inducing periodic variability \citep{peng_warped_2024}. This warp may develop during the accretion of surrounding gas onto the disk or arise from the torque exerted during close encounters with a black hole companion, approaching from a non-coplanar direction \citep{papaloizou_dynamics_1995}, from non-coplanar direction.

    \item[3.] This scenario depicts a TDE characterized by significant variability in the X-ray light curve, whereas the optical light curve exhibits smaller, inversely correlated variations. We designate this case as the ``\textit{Blinker}''. This behavior resembles J0456-20, whose overall X-ray light curve adheres to a $-5/3$ power-law decay, typical of a TDE, and presents several rapid rise-and-decay outbursts approximately 200 days apart. Similarly, its UV light curve displays a gradual decline accompanied by slight periodic fluctuations. Notably, the UV and X-ray light curves exhibit an inverse relationship, where dips in the UV light curve coincide with peaks in the X-ray light curve. We will dive deeper into modeling this source in Section \ref{subsec:Application}.
   
    \item[4.] In this scenario, the viewing angle evolves over a wide range, even crossing the $90^{\degree}$, leading to repeated transitions from optical-dominated ($\theta \sim 90^{\degree}$) to X-ray-dominated emission. Notably, it has two optical peaks in a precession cycle. This occurs because the tilted angle is exceptionally large, enabling the observer to observe the disk in an edge-on direction twice during a single precession cycle. Given that the disk is observed edge-on at certain times, this type of precessing disk ought to be optical-dominated during those intervals. Subsequently, as the optical emission diminishes, the system could transition to being X-ray bright. The modulation pattern of this case is similar to that of the ``\textit{Blinker}'', but an `abac' pattern where the two X-ray peaks differ within a precession cycle. Thus, we refer to it as the ``\textit{Siren}''. However, to the best of our knowledge, no such TDE has been observed to date. This absence could imply the presence of potential selection biases or some physical reasons, as discussed in Section \ref{subsec:Application}.

    \item[5.] When the precession period exceeds the mass fallback timescale, that is, when $T_{\rm pre} = 2\pi/\omega \gtrsim t_{\rm fb}$, we cannot see the periodic variability in the light curve. Instead, we may observe an X-ray brightening at late times as the viewing angle aligns with the near-polar direction of the disk as a result of the precession effect. This phenomenon could possibly explain the sharp late-time rise in X-rays observed in certain TDEs, such as ASASSN-15oi \citep{hajela_eight_2025}, AT 2019azh \citep{Liu_AT2019azh_2022}, and OGLE16aaa \citep{kajava_rapid_2020}. However, it is important to note that late-time X-rays in TDEs have also been attributed to delayed accretion or outflow-torus interactions \citep{zhuang_late_2021, mou_years-delayed_2021}. If the late-time X-ray emission stems from the precession effect, we might anticipate a second X-ray flare at a later stage. Conversely, if the alignment process occurs more rapidly than the precession, we may not observe this secondary flare. Similarly, if the disk's polar direction is initially facing the observer, early X-ray emission could be detected at the very beginning, but it would soon diminish as the disk shifts to other directions due to slow precession. If the polar direction subsequently aligns with the observer again, a late-time X-ray rebrightening would be observed. A TDE candidate exhibits very early X-ray emission, AT 2022dsb \citep{malyali_transient_2024}, may belong to this case. % The evolution of polarization could serve as an additional signature for the precession model, potentially aiding in future model testing. %If the polar direction in the early time, we may see a X-ray bright, but soon disappear as the viewing angle far away late time.
\end{enumerate}

The precession signal can be decomposed into rapid and slow scenarios, with the former producing periodic modulations and the latter enabling detection of late-time X-ray transitions. Cases 1--4 correspond to rapid precession and are illustrated in Figure \ref{fig:pattern} and summarized in Table \ref{tab:patterns}.

It is noteworthy that as the accretion rate declines to a sub-Eddington state in later stages, the disk transitions into a geometrically thin disk. The viscous torque induces rapid alignment of the tilted disk, as documented in previous studies \citep{bardeen_lense-thirring_1975, nelson_hydrodynamic_1999, stone_observing_2012}. We give a detailed calculation on the alignment timescale in appendix \ref{sec:alignment}. Alternatively, it may tear the disk apart, leading to a fast precession of the inner portion and subsequent alignment, while the outer part remains tiled. Therefore, we would not detect the light curve modulation in late time of TDEs. The corresponding timescale of Super-Eddington phase is given by \citep{Rees_Tidal_1988,Chen_Tidal_2018}
\begin{equation}
    t_{\rm Edd} \simeq 800\ \left(\frac{\eta_{\rm NT}}{0.12}\right)^{3/5} M_6^{-2/5} r_*^{3/5} m_*^{1/5}\ {\rm day},
\end{equation}
Since $t \lesssim  t_{\rm Edd}$ in Figure \ref{fig:Lx_Lop_dec}, the approximation used here is reasonable.
%%%%%%%%%%%%%%%%%%%%%%%%%%%%%%%%%%%%%%%%

%%%%%%%%%%%%%%%%%%%%%%%%%%%%%%%%%%%%%%%%
\section{Discussion}
\label{sec:discussion}
%%%%%%%%%%%%%%%%%%%%%%%%%%%%%%%%%%%%%%%%%
\subsection{Unexplored Dynamics and Radiative Signatures of Realistic Tilted Disks}
\label{subsec:realistic}
In this paper, we impose an artificial global rigid-body precession to explore the simply dynamics and radiative signatures of a precession disk in TDE. Actually, simulations conducted by \cite{fragile_global_2007} and \cite{liska_formation_2018} have revealed the global precession of a thick accretion disk. Given that the warp induced by the LT effect can propagate rapidly in a wave-like manner within a disk, particularly in the case of a small disk formed during a TDE, it is reasonable to expect global rigid-body precession even for large tilt angles. It is worth noting that some simulations of thin disks with large tilt angles suggest that such disks cannot precess as a rigid body; instead, they may undergo tearing due to differential precession, wherein the outer disk fails to precess coherently while the inner disk rapidly aligns via the Bardeen-Petterson effect \citep{liska_h-amr_2022,musoke_disc_2023}. As the accretion rate decreases to lower levels in the late stages of a TDE, we anticipate that the disk will become thin and could potentially experience tearing.

The radiative characteristics of a realistic precessing disk experiencing a declining accretion rate would exhibit subtle deviations from our simplified calculations presented here. In our basic analysis, we assumed a constant viewing angle dependency (as depicted in Figure \ref{fig:spectrum}) to quantify the effects of precession. However, in reality, higher accretion rates intensify disk winds, causing a shift in emission towards the optical/UV bands. It is imperative to consider time-dependent spectral evolution, especially when significant variations in the accretion rate occur. Changes in the spectrum induced by the accretion rate can also impact the viewing angle dependency of both X-ray and optical/UV luminosities (as shown in Figure \ref{fig:Lratio_angle}). For lower accretion rates, the range of viewing angles for the Blinker and the Siren may shift towards larger angles. 

Furthermore, the spectral output derived from our current analysis is  restricted by the quasi-1D calculation performed using the Sedona code. \cite{parkinson_multidimensional_2025} employed a 2.5D radiative transfer code for post-processing the simulations and discovered that the spectrum may not exhibit as strong a dependence on the viewing angle as previously thought, since photons may preferentially escape in polar directions where the optical depth is lower. Therefore, a more comprehensive and detailed investigation into the dynamics and radiative signatures of precessing disks in TDEs is warranted for future research endeavors.

\subsection{Where are the Blinkers?}
\label{subsec:Application}
The recently reported nuclear transient eRASSt J045650.3-203750  \citep[hereafter, J0456-20, ][]{Liu_J045650_2023,liu_rapid_2024}, identified by extended ROentgen Survey with an Imaging Telescope Array (eROSITA) in a quiescent galaxy at redshift of $z \sim 0.077$, exhibits long-term decay with superimposed short-term quasi-periodic X-ray variability. We suggest that this behavior may be due to a precessing thick disk with a declining accretion rate.

\begin{figure*}
\centering
 \includegraphics[scale=0.5]{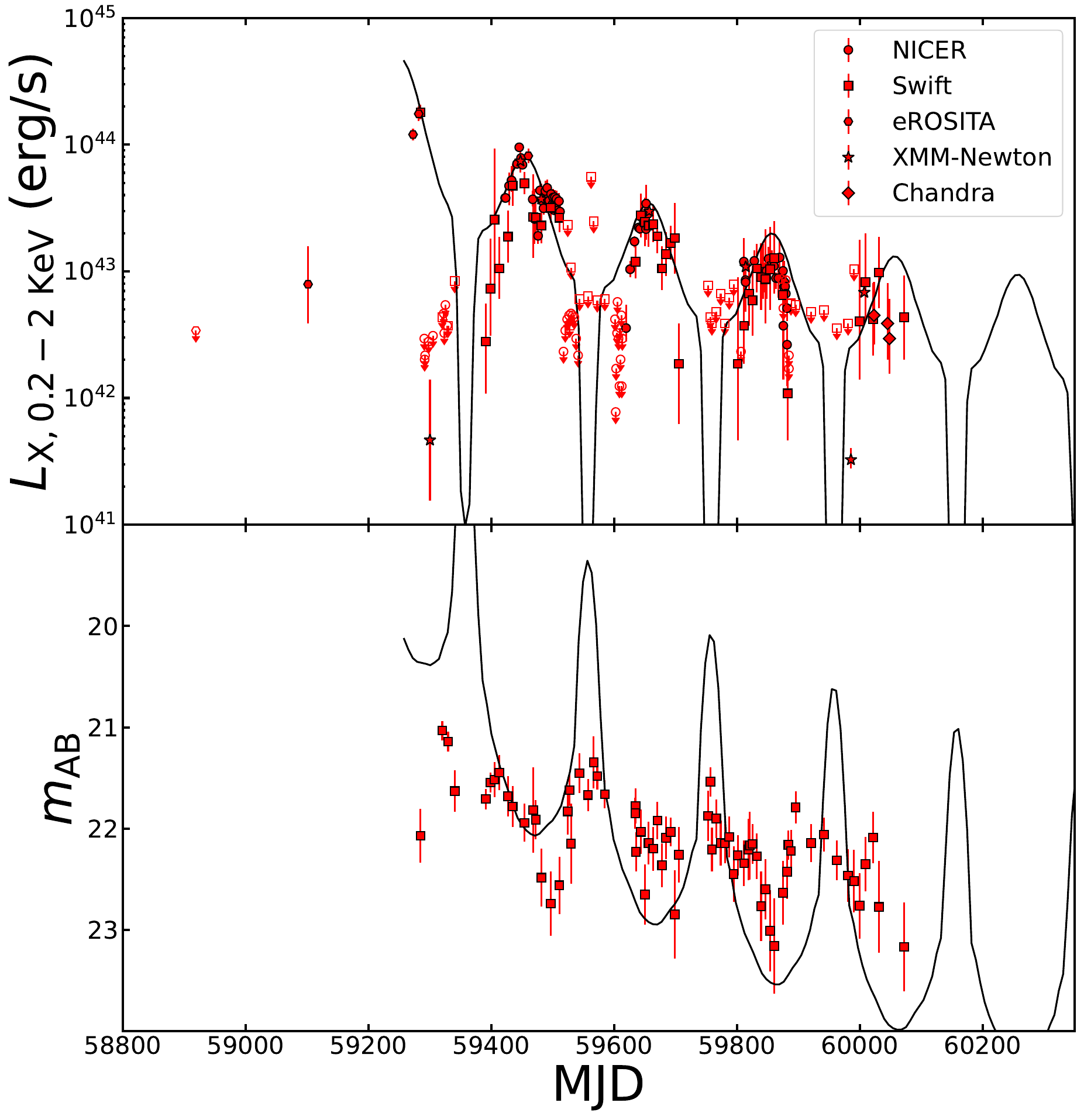}
\caption{X-ray and NUV data of J0456-20 on top of the modeled light curves. The model parameters are $t_{\rm fb} \simeq 108$ day, $t_0 \simeq 59258$ MJD, $L_0 \simeq 63$, $\varphi \simeq 12^{\degree}(40.3^{\degree})$, $\theta \simeq 40.3^{\degree}(12^{\degree})$. The corresponding viewing angle $i(t)$ evolves from $28^{\degree}$ to $52^{\degree}$.}
\label{fig:LC_fit}
\end{figure*}

The X-ray and UV light curves of J0456-20 are shown in Figure \ref{fig:LC_fit}, where the X-ray data are obtained from \cite{liu_rapid_2024}. In the calculation of the observed luminosity, we adopt a flat {$\Lambda$}CDM cosmology with $H_0 = 67.7\ {\rm km\ s^{-1}\ Mpc^{-1}}$ and $\Omega_m = 0.308$ \citep{planck_collaboration_planck_2020}.

The overall X-ray light curve exhibits a $-5/3$ power-law decay, consistent with the characteristic behavior of a typical TDE. However, it also displays several rapid rise-and-decay outbursts, separated by about 200 days, which deviate from the standard TDE profile. Similarly, the NUV light curve shows a gradual decay with slight periodic variations. Notably, the NUV and X-ray light curves show an inverse relationship: dips in the NUV light curve coincide with the peak in the X-ray light curve.

These observational characteristics are consistent with the ``\textit{Blinker}'' as discussed in Section \ref{subsec:LC_Macc}. Therefore, we propose that it arises from emissions originating in a precessing tilted disk, formed as a result of the TDE.

The model light curve is shown in Figure \ref{fig:LC_fit}, with parameters $t_{\rm fb} \simeq 108$ days, $t_0 \simeq 59258$ MJD, $L_0 \simeq 63$, $\varphi \simeq 12^{\degree}(40.3^{\degree})$, $\theta \simeq 40.3^{\degree}(12^{\degree})$. The results indicate a precession period of about $200$ days. The viewing angle evolves from $ 28^{\degree}$ to $52^{\degree}$ in each precession cycle.

Furthermore, the overall light curve decay timescale $t_{\rm fb} \sim 108$ days is approximated to the fallback timescale of a TDE, i.e., $t_{\rm fb} \simeq 130 (M_{\rm h}/10^7M_{\odot})^{1/2} (R_*/R_{\odot})^{3/2} (M_*/M_{\odot})^{-1}$ days \citep{Rees_Tidal_1988}, where $M_*$ and $R_*$ are the stellar mass and radius, respectively. Here we adopt the $10^7\ M_{\odot}$ SMBH mass, which is estimated by the $M-\sigma$ relation \citep{Liu_J045650_2023}.

%Additionally, we can constrain the SMBH's spin of J0456-20. In Figure \ref{fig:constrain_a}, we find that a spin parameter $|a| \gtrsim 0.1$ is required to explain the $200$-day precession period with $10^7\ M_{\odot}$ SMBH's mass. This result aligns with observations of SMBHs' spin in AGNs \citep{piotrovich_estimate_2023}, which suggest that many SMBHs are rapidly or moderately rotating.

Notice that, the model light curve presented here is not a precise fit to the observed data; rather, it serves to offer a qualitative explanation of the observed light curve behavior. The model's tendency to overestimation of NUV flux likely arises from many factors. For instance, the assumption of a fixed spectral dependence overlooks the impact of disk wind density and accretion rate variations on optical/UV reprocessing. Additionally, the observed luminosities may be subject to extinction effects. Furthermore, as indicated in Section \ref{subsec:realistic}, a more accurate spectrum would necessitate calculations via multi-dimensional radiative transfer, as demonstrated by \cite{parkinson_multidimensional_2025}.  As this work focuses on qualitative estimates, we defer precise fitting to future studies incorporating wind models and extinction corrections.

In conclusion, the precessing TDE disk model provides a comprehensive interpretation of the multiwavelength behavior of J0456-20. However, the scarcity of detected \textit{Blinker} candidates raises the critical question: What fraction of TDEs host precessing disks with significant tilt angles?

If the disk tilt angle $\theta$ and the mutual inclination $\varphi$ between the observer's line of sight and the SMBH's spin direction are randomly oriented in space, the most probable configuration would be $\theta \sim \varphi \sim 90^{\degree}$ (i.e., both the disk and observer's line of sight are perpendicular to the SMBH's spin). In this scenario, we would expect most TDEs to exhibit strong modulation due to precession. However, this contradicts observations, which show that the majority of TDEs display smooth light curves or only weak modulation. This discrepancy may arise from intrinsic physical factors or observational selection biases. 

There are several unconsidered physical effects. Stars may preferentially disrupt in the SMBH's equatorial plane, aligning the disk with the SMBH spin axis. For example, axisymmetric nuclear stellar clusters (e.g. saucer-like structures from galaxy mergers) could confine disruptions to the equatorial plane \citep{zhong_supermassive_2015}. Additionally, a pre-existing accretion disk surrounding the SMBH may capture stars via dynamical friction, enhancing TDE rates and further biasing disruptions toward the equatorial plane \citep{wang_changing-look_2024}. Another physical mechanism arises from the inefficient formation of disks in TDEs. If a star's orbit is significantly tilted relative to the SMBH's equatorial plane, the returning debris stream may fail to intersect, preventing stream-stream collisions. This inefficient disk formation leads to fainter TDEs \citep{dai_impact_2013, Guillochon_A_2015, Jiang_PROMPT_2016, jankovic_spin-induced_2024}.

Despite these potential suppression effects, the light curves of \textit{Blinker} and \textit{Siren} would still differ markedly from typical TDEs, potentially causing them to be overlooked in surveys. Future observational campaigns with improved sensitivity and cadence are essential to detect and characterize these rare systems.

\subsection{Precession: Rapid or Slow?}
\label{subsec:precession_timescale}
In Section \ref{subsec:LC_Macc}, we conducted a comprehensive analysis of two distinct precession regimes in TDE accretion disks, rapid precession characterized by periodic modulation in the light curve, and slow precession, which manifests itself as gradual, secular changes in X-ray emission.

The precession timescales are governed by the interplay of critical system parameters, including the SMBH's spin, disk density, and disk size. The detailed calculation is given in Appendix \ref{appendix:precession_timescale}.

For a rapidly spinning SMBH and a compact, mass-concentrated accretion disk, the LT torque induces rapid precession, with timescales as short as minutes to days. Conversely, in systems with a slowly rotating SMBH, or an extended, outer-mass-dominated disk, precession is weakened, leading to slower evolution over months to years.

\subsection{Jet Formation and Precession in TDEs}
\label{subsec:jet}
A super-Eddington accretion disk in a TDE can launch a relativistic jet powered by either the Blandford-Znajek (BZ) mechanism \citep{blandford_electromagnetic_1977}, which extracts rotational energy from the spinning SMBH, or the Blandford-Payne (BP) mechanism \citep{blandford_hydromagnetic_1982}, which taps into the disk's rotational energy. BZ-powered jets are expected to align with the SMBH's spin axis, as they originate from the ergosphere where frame-dragging effects dominate. Thus, in the absence of disk-jet coupling, such jets should not precess. BP-powered jets, however, are anchored to the accretion disk and would naturally precess if the disk itself is warped or misaligned with the SMBH spin. Interestingly, some simulations suggest that even BZ-powered jets can align with the disk axis \citep{liska_formation_2018, liska_phase_2023, ressler_wind-fed_2023}, potentially due to strong disk outflows exerting torques that reorient the jet. The exact alignment mechanism remains debated.

There is some observational evidence for jet precession. The nearby active galaxy M87 exhibits co-precession of its jet and compact accretion disk \citep{cui_precessing_2023, cui_co-precession_2025}. Additionally, variability in some GRB light curves has been attributed to jet precession \citep{lei_model_2007, liu_jet_2010, zhang_grb_2023}.
Swift J1644+57, a strong evidence of jetted TDE, discovered by Swift \citep{Burrows_Relativistic_2011}, showed X-ray variability interpreted as jet precession \citep{lei_black_2011, lei_frame_2012,}. EP250702a, detected by the Einstein Probe \citep{GCN40906}, this fast X-ray transient is spatially and temporally associated with multiple GRBs (GRB250702D, B, E; \citealt{GCN40891, GCN40931}). Its light curve and multiwavelength spectrum suggest a jet-powered TDE with an intermediate-mass black hole, where multiple GRBs may arise from precession \citep{Levan2025arXiv}.

Despite these cases, most TDEs show no clear jet signatures, likely due to intrinsically only a small fraction of TDEs producing relativistic jets, narrow jet opening angles may miss our line of sight, or the jet could be choked by the disk outflow, as the jet precesses \citep{teboul_unified_2023,lu_misaligned_2024,yuan_propagation_2025}. According to \cite{teboul_unified_2023}, if precession occurs rapidly enough, disk winds can envelop the system on a large scale. Consequently, the precessing jet may or may not break free from the wind region during its early stages. For instance, the highly variable X-ray emissions observed in the early phase of Swift J1644+57 could be attributed to a successfully escaping precessing jet.

%%%%%%%%%%%%%%%%%%%%%%%%%%%%%%%%%%%%%%%%
\section{Conclusion}
\label{sec:conclusion}
In this paper, we model the radiative properties of a super-Eddington disk which is misaligned with the black hole spinning axis and therefore undergoing LT precession. The results are applied to explain and predict multiwavelength light curve or variation behavior of TDEs. Our key findings are as follows:
\begin{enumerate}
\item A super-Eddington accretion disk naturally produces both X-ray emissions at polar directions and 
reprocessed optical/UV emissions at moderate to large inclinations, as previously demonstrated \citep{dai_unified_2018}. We find that the UV/optical fluxes remain rather insensitive to the viewing angle for a wide range of inclinations, while the X-ray emission varies faster with the viewing angle. As a result, as the disk undergoes precession, the X-ray light curve will display a more significant amplitude of modulation than its optical light curve. %In our primary model, we make the straightforward assumption that the viewing angle to the disk changes periodically. However, in a more realistic scenario, during precession, the outflow can periodically occupy the polar region. Therefore, it is reasonable to hypothesize that this could result in less variability in the optical/UV light curve than our current simplified model calculations indicate.
    
\item The modulation pattern and amplitude of the multi-wavelength light curves during a precession cycle are governed by the variation of the viewing angle with respect to the disk, which is in turn governed by two key parameters: the disk tilt angle $\theta$ relative to the spin axis of the SMBH, and the inclination $\varphi$ between the observer's line of sight and the SMBH's spin direction. The viewing angle range during a precession cycle is described by Equation (\ref{eq:irange}). 

\item Different combinations of  $\theta$ and $\varphi$ lead to three main precession patterns, as illustrated in Figure \ref{fig:pattern} and summarized in Table \ref{tab:patterns}. We give a brief summary as follows.

\begin{enumerate}

\item When either $\theta =0^{\degree}$ or $\varphi =0^{\degree}$, the observer will detect a TDE without precession, consistent with TDE candidates showing no fluctuations, which we classify as ``\textit{Smooth-TDEs}''.

\item The \textit{Dimmer}: When the viewing angle range falls within the outflow-cleared funnel in the polar direction or the thick outflow in the equatorial direction, it results in minor fluctuations in the light curve in the X-ray and optical/NUV bands, respectively. This specific scenario corresponds to TDE candidates that show little variability, such as ASASSN-14li.

\item The \textit{Blinker}: When the viewing angle spans a moderate range and shifts between the thick outflow region and the polar region, it induces intense variability in the X-ray light curve. X-ray emission may disappear when the light of sight is directed towards the thick outflow region, while optical and NUV emissions have a slight increase. This inverse relationship between the X-ray and optical/UV light curves aligns with observations of the TDE candidate J0456-20 \citep{Liu_J045650_2023,liu_rapid_2024}. 

\item The \textit{Siren}: For extremely large tilt angles and $\varphi$, the viewing angle can span a range from  $\sim 0^{\circ}$ to $180^{\circ}$. This sweeping motion carries the observer's line of sight from one side of the polar direction to the opposite side, resulting in extreme large modulations in both optical/UV and X-ray light curves. During a single precession cycle, both optical and X-ray emissions are observed to exhibit two peaks. Moreover, the two X-ray peaks can be distinct in flux and other characteristics.

\end{enumerate}

\end{enumerate}

Our work provides a valuable tool for modeling the multiwavelength emission properties of TDEs with precessing disks, through which our understanding of the geometry and dynamics of precessing thick disks can be greatly advanced. 
Furthermore, detecting and analyzing precession-induced modulations in the light curves of TDEs can provide crucial constraints on two key physical parameters: the viewing angle of the accretion disk relative to the observer and the tilt angle of the disk with respect to the spin axis of the SMBH. Last but not least, the properties of the precession-induced modulation can offer promise in constraining the spin of SMBHs.

With the advent of upcoming and newly operational time-domain telescopes conducting sensitive all-sky surveys, we anticipate detecting more TDEs with precession signals, providing deeper insights into the underlying physics of TDEs and precessing disks, and testing principles of general relativity. For instance, the Einstein Probe X-ray telescope can discover numerous TDEs and conduct follow-up observations in the soft X-ray band. The Imaging X-ray Polarimetry Explorer (IXPE) can detect X-ray polarization, offering a powerful tool for studying precessing disks in TDEs. The upcoming optical telescopes, such as the China Space Station Telescope (CSST) and the Legacy Survey of Space and Time (LSST) by the Vera C. Rubin Observatory, are expected to discover hundreds to thousands of TDEs per year in the optical band \citep{bricman_prospects_2020}, significantly expanding the TDE sample. By combining X-ray and optical telescopes, one can observe TDE emission across multi-wavelengths and constrain their precession patterns, using which one can test precessing disk theories  and constrain SMBH spin properties.

%%%%%%%%%%%%%%%%%%%%%%%%%%%%%%
\appendix
\counterwithin{figure}{section}

\twocolumngrid

\section{Spectra of a super-
Eddington disk in different inclinations}

\begin{figure}		
\centering
 \includegraphics[scale=0.5]{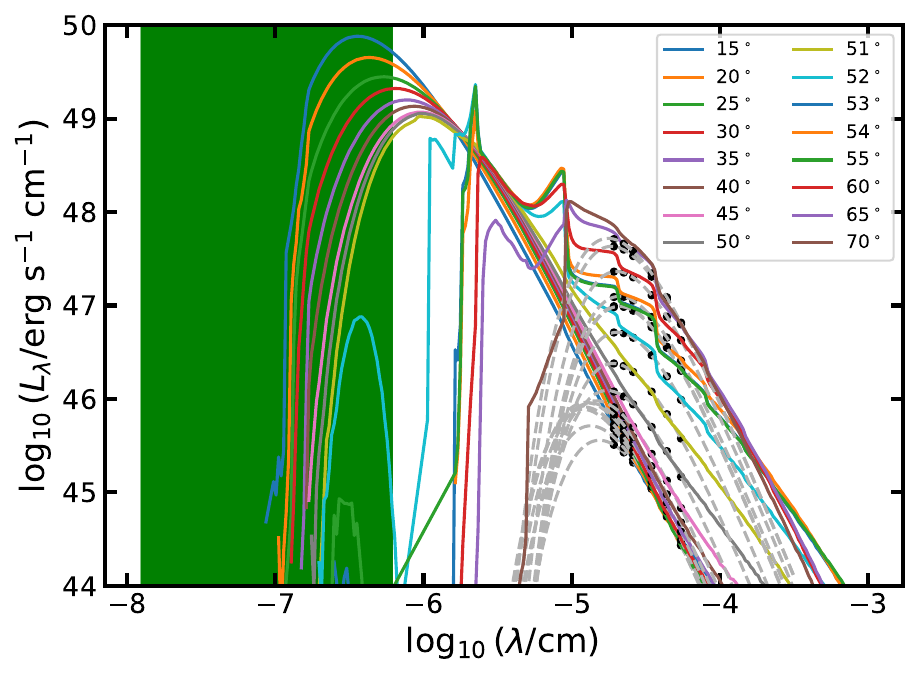}
    \caption{Escaped continuum emission spectra of a super-Eddington disk with an accretion rate of $\dot{M}_{\rm acc}=12\ \dot{M}_{\rm Edd}$. The different colors represent various viewing angle. The green shade area highlights the $0.2 - 10$ keV X-ray band, while the black points denote $L_{\lambda}$ for the \textit{Swift}/UVOT bands (v, b, u, UVW1, UVM2 and UVW2 filters). The dashed gray lines indicates the blackbody fitting in optical/NUV bands.}
    \label{fig:spectrum}
\end{figure}

\section{Precession timescale}
\label{appendix:precession_timescale}
The combined LT torques across different regions of the disk drive its precession. The precession period depends on the density of the disk surface ($\Sigma(R) \propto R^{-\zeta}$), the inner and outer radii of the precessing disk ($R_{\rm in}$ and $R_{\rm out}$), as well as the mass and spin of the SMBH \citep{stone_observing_2012,Chen_AT2019avd_2022}. It can be expressed as:
\begin{equation} \label{eq:T_pre}
    \begin{split}
    T_{\rm pre} &\simeq \frac{\pi c^3}{|a|G^2M_{\rm h}^2} \frac{\int^{R_{\rm out}}_{R_{\rm in}} R'^{3/2} \Sigma(R')\ dR'}{\int^{R_{\rm out}}_{R_{\rm in}} R'^{-3/2} \Sigma(R')\ dR'} \\
    &= T_{\rm LT} g(r_{\rm out}, \zeta),
    \end{split}
\end{equation}
where 
\begin{equation} \label{eq:T_LT}
    T_{\rm LT} = \frac{\pi c^3}{|a|G^2M_{\rm h}^2} R_{\rm in}^3
\end{equation}
represents the LT precession period for a gas particle located in the innermost circular orbit $R_{\rm in}$. Here, $M_{\rm h}$ denotes the SMBH's mass, and $a$ is the dimensionless spin parameter. The function 
\begin{equation} \label{eq:g_pre}
    g(r_{\rm out}, \zeta) = \frac{\int^{r_{\rm out}}_{r_{\rm in}} r'^{3/2} \Sigma(r')\ dr'}{\int^{r_{\rm out}}_{r_{\rm in}} r'^{-3/2} \Sigma(r')\ dr'}
\end{equation}
depends on the ratio $r_{\rm out} \equiv R_{\rm out}/R_{\rm in} \gg 1$ and the power-law index of the surface density $\zeta$.

When $\zeta \gtrsim 5/2$, the function $g(r_{\rm out}, \zeta)$ approaches unity. In this case, the inner part of the disk contributes the majority of the LT torques responsible for the precession, resulting in rapid precession with a period $T_{\rm pre} \sim T_{\rm LT}$. In contrast, if $\zeta \lesssim -1/2$, then $g(r_{\rm out},\zeta) \simeq r_{\rm out}^3$. Here, the outer part of the disk contributes most of the LT torques, leading to slow precession with a period $T_{\rm pre} \sim T_{\rm LT} r_{\rm out}^3$.

The inner radius of the disk, $R_{\rm in}$, can be approximated as the innermost stable circular orbit, which depends on the spin of the SMBH. For a prograde ($a>0$) or retrograde ($a<0$) orbit, it is given by:
\begin{equation} \label{eq:Rin}
    R_{\rm in} \simeq \left(3+Z_2 - \frac{a}{|a|}\sqrt{(3-Z_1)(3+Z_1+2Z_2)}\right) R_{\rm g},
\end{equation}
where $Z_1 = 1+(1-a^2)^{1/3}[(1+a)^{1/3}+(1-a)^{1/3}]$ and $Z_2 = (3a^2+Z_1^2)^{1/2}$. For prograde and retrograde orbits with the maximum spin $|a|=1$, $R_{\rm in} = 1$, $9 R_{\rm g}$, respectively, where $R_{\rm g} = GM_{\rm h}/c^2$ is the gravitational radius.

%The surface density of a super-Eddington accretion disk follows $\Sigma \propto R^{-\zeta}$, where $\zeta \simeq -3$ to $-2$ \citep{jiang_super-eddington_2019}. As a result, the precession torque is primarily determined by the outer region of the disk, and $T_{\rm pre}$ can be expressed as \citep{Chen_AT2019avd_2022}:

For a newly formed TDE disk, $R_{\rm out}$ is expected to be close to the circularization radius, roughly $2R_{\rm T}$, while $R_{\rm T} = R_*(M_{\rm h}/M_*)^{1/3}$ being the tidal radius. Consequently, we obtain $r_{\rm out} \simeq 10$ and $10^2 (M_{\rm h}/10^6M_{\odot})^{-2/3} (R_*/R_{\odot}) (M_*/M_{\odot})^{-1/3}$ for prograde and retrograde orbits, respectively. 

If the inner part of the disk governs the LT precession, the precession occurs rapidly with a period $T_{\rm pre} \simeq T_{\rm LT} \simeq 10$ to $10^3 |a|^{-1} (M_{\rm h}/10^6M_{\odot})\ {\rm s}$. The lower and upper values correspond to the prograde and retrograde orbits, respectively. This could potentially explain some QPO-like signals observed in TDEs with short periods. In contrast, if the outer part of the disk dominates the LT precession, the precession is slow, with a period $r_{\rm out}$ times that of the previous one, spanning timescales from days to years.

\section{Alignment timescale}
\label{sec:alignment}
Typically, the precession of a thick disk can persist for a long time, continuing until the disk becomes geometrically thin. Here, we examine two mechanisms that drive the alignment process, as outlined in \cite{franchini_lense-thirring_2016}. The first mechanism involves viscous shear, which dissipates the disk's precession energy at a rate proportional to $\alpha$, the dimensionless viscosity parameter. The corresponding alignment timescale is given by
\begin{equation} \label{eq:t_Bate}
    t_{\rm Bate} \simeq \alpha^{-1} \left(\frac{H}{R}\right)^2 \frac{\Omega}{\omega^2}.
\end{equation}
Here, $H/R$ represents the relative scale height, and $\Omega = (GM_{\rm h}/R^3)^{-1/2}$ is the Keplerian angular velocity at the outer radius $R_{\rm out}$, as detailed in \cite{bate_observational_2000}. 

The second mechanism comes into play as the accretion rate decreases. The disk gradually cools to $\alpha \gtrsim H/R$,  at which point the propagation of the warp within the disk becomes diffusive. This leads to rapid damping of any warps and subsequent alignment of the disk, as discussed in \cite{bardeen_lense-thirring_1975} and \cite{nelson_hydrodynamic_1999}. For a thick disk, which is typical in TDEs, $\alpha \lesssim H/R$ initially holds true. Consequently, warp propagation is wave-like, resulting in rigid precession of the disk. This scenario was considered in \cite{stone_observing_2012}, where the alignment timescale was calculated as:
\begin{equation} \label{eq:t_SL}
    t_{\rm SL} \simeq t_{\rm fb} \left(15\frac{f}{X}\frac{\dot M_{\rm peak}}{\dot M_{\rm Edd}} \frac{R_{\rm g}}{R_{\rm out}}\right)^{3/5}.
\end{equation}
In this equation, $f=1-(R_{\rm in}/R)^{1/2}$ and $X = \alpha/(1-8\alpha^2/3/f)$. $\dot M_{\rm peak} \simeq M_*/3t_{\rm fb}$ is the peak mass fallback rate with a corresponding timescale $t_{\rm fb}$. $\dot{M}_{\rm Edd}$ is the Eddington mass accretion rate defined as $\dot{M}_{\rm Edd} \equiv L_{\rm Edd}/(\eta_{\rm NT} c^2)$, where $L_{\rm Edd}$ is the Eddington luminosity, and $\eta_{\rm NT} \simeq 0.12$.

\begin{figure*}
\centering
\begin{minipage}[t]{0.45\textwidth}
\centering
 \includegraphics[scale=0.5]{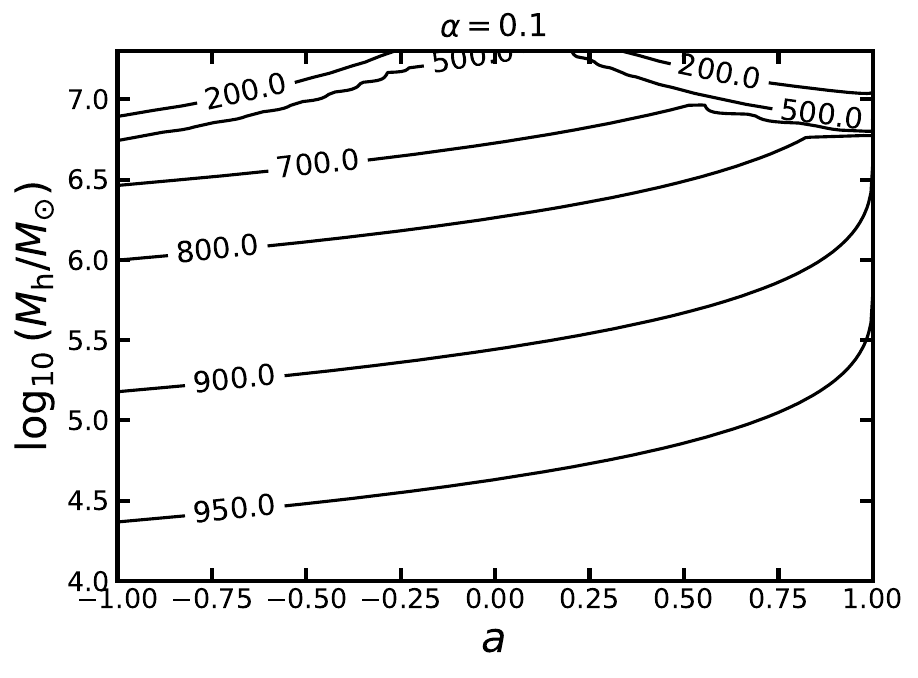}
 \end{minipage}
\begin{minipage}[t]{0.45\textwidth}
\centering
 \includegraphics[scale=0.5]{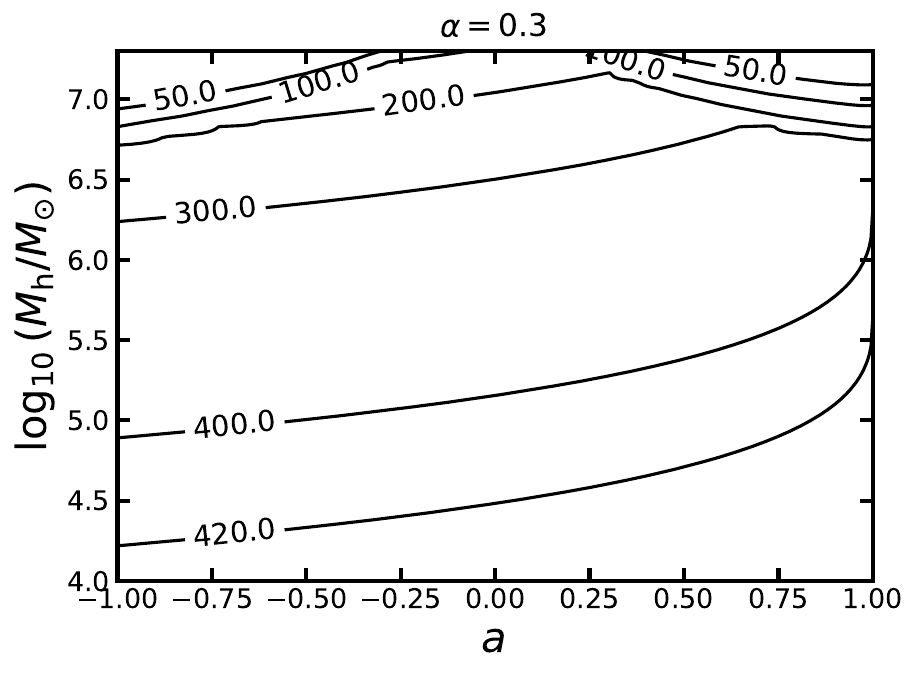}
 \end{minipage}
    \caption{Alignment timescale of a precessing disk, formed by the disruption of a Sun-like star by a SMBH with different mass and spin. The left and right panels represent the results with $\alpha = 0.1$ and $\alpha=0.3$, respectively. The contour lines indicate the alignment timescale $t_{\rm align}$ in unit of day.}
\label{fig:t_align}
\end{figure*}

Combining these two mechanisms, the alignment timescale can be estimated as
\begin{equation} \label{eq:t_align}
    t_{\rm align} = \min(t_{\rm Bate}, t_{\rm SL}).
\end{equation}
Assuming the outer radius of the precessing disk is $R_{\rm out} \simeq 2 R_{\rm T}$ and that the disrupted star is a Sun-like star, Figure \ref{fig:t_align} presents the alignment timescale for different SMBH mass and spin.

Generally, disk cooling is the main driver of alignment, resulting in $t_{\rm align} \simeq t_{\rm SL} \propto \alpha^{-3/5}$. The corresponding timescale typically ranges from one to several years. If the SMBH's mass is $\gtrsim 10^7 M_{\odot}$ and its spin is $|a| \gtrsim 0.2$, the viscous dissipation becomes efficient, leading to $t_{\rm align} \simeq t_{\rm Bate} \propto \alpha^{-1}$ and a faster alignment timescale of a few months. However, in reality, the inflowing debris stream can introduce additional misaligned angular momentum into the disk upon joining, potentially prolonging the alignment timescale.

The TDE candidate AT2020ocn is believed to have a precessing disk. Its quasi-periodic X-ray variability is associated with changes in viewing angle, exhibiting a 10-day period \citep{cao_tidal_2024}. More intriguingly, its variability displays an overall decreasing trend over approximately 200 days, which implies the alignment process.

Magnetic forces also play a crucial role in disk alignment. Some studies demonstrate that magnetically arrested disks (MADs) are compelled to align with the black hole's spin axis due to the saturation of magnetic flux in the black hole's ergosphere and inner accretion disk. As the black hole rotates, it twists this saturated magnetic flux. In the MAD state, the magnetic field exerts a dominant influence on the dynamics of the inner disk. Consequently, the inner disk adjusts its orientation in accordance with the alignment of the magnetic field \citep{mckinney_alignment_2013,ressler_wind-fed_2023,chatterjee_misaligned_2023,fragile_tilted_2024}. However, the efficiency of this forced alignment depends on the strength and radial extent of the magnetic saturation. Currently, several key questions remain unanswered. For instance, it is uncertain whether the magnetic field strength in the thick disk formed during a TDE is adequate to maintain a MAD state. Additionally, the precise level of magnetic flux required for rapid alignment remains an open question.

\begin{acknowledgments}
This work is supported by the National Natural Science Foundation of China and the Hong Kong Research Grants Council (the NSFC Type C Youth Project 12503053, N\_HKU782/23, HKU 109001456). We also acknowledge the useful discussion with Hongxuan Jiang, Benny Trakhtenbrot, Bing Zhang, Rongfeng Shen, Yanan Wang, Lars Lund Thomsen. We thank the participants of the TDE FORUM (Full-process Orbital to Radiative Unified Modeling) online seminar series for their inspiring discussions.
\end{acknowledgments}

%%%%%%%%%%%%%%%%%%%%%%%%%%%%%%%%%%%%%%%%%
%\begin{contribution}
%\end{contribution}
%%%%%%%%%%%%%%%%%%%%%%%%%%%%%%%%%%%%%%%%%%%%%%%%%%%%%
\bibliography{cited}{}
\bibliographystyle{aasjournalv7}

\end{CJK*}
\end{document}